\documentclass[twocolumn,prx,showpacs,superscriptaddress,10pt]{revtex4-1}
\usepackage{graphicx}
\usepackage{amsmath}
\usepackage{amsfonts}
\usepackage[table]{xcolor}
\usepackage[caption=false]{subfig}
\usepackage{array}
\usepackage{floatrow}
\usepackage{pbox}
\usepackage[colorlinks=true,linkcolor=blue,citecolor=blue,filecolor=cyan,urlcolor=blue,breaklinks=true]{hyperref}

\newcommand{\mrd}{\mathrm d}
\newcommand{\mre}{\mathrm e}

\newcommand{\mean}[1]{\left\langle #1 \right\rangle}
\newcommand{\ord}[1]{\mathcal{O}\left(#1\right)}
\newcommand{\pstat}{p^\mathrm{s}}

\newcommand{\sm}{s_\mathrm{m}}
\newcommand{\mep}{\sigma}  
\renewcommand{\vec}{\boldsymbol}

\begin{document}
\title{Universal bounds on current fluctuations}

\author{Patrick Pietzonka}
\affiliation{ II. Institut f\"ur Theoretische Physik, Universit\"at Stuttgart,
  70550 Stuttgart, Germany}
\author{Andre C. Barato}
\affiliation{Max Planck Institute for the Physics of Complex Systems,
  N\"othnitzer Stra\ss{}e 38, 01187 Dresden, Germany}
\author{Udo Seifert}
\affiliation{ II. Institut f\"ur Theoretische Physik, Universit\"at Stuttgart,
  70550 Stuttgart, Germany}

\parskip 1mm

\begin{abstract}

  For current fluctuations in non-equilibrium steady states of Markovian
  processes, we derive four different universal bounds valid beyond the
  Gaussian regime. Different variants of these bounds apply to either the
  entropy change or any individual current, e.g., the rate of substrate
  consumption in a chemical reaction or the electron current in an electronic
  device. The bounds vary with respect to their degree of universality and
  tightness.  A universal parabolic bound on the generating function of an
  arbitrary current depends solely on the average entropy production.  A
  second, stronger bound requires knowledge both of the thermodynamic forces
  that drive the system and of the topology of the network of states. These
  two bounds are conjectures based on extensive numerics. An exponential bound
  that depends only on the average entropy production and the average number
  of transitions per time is rigorously proved. This bound has no obvious
  relation to the parabolic bound but it is typically tighter further away
  from equilibrium. An asymptotic bound that depends on the specific
  transition rates and becomes tight for large fluctuations is also
  derived. This bound allows for the prediction of the asymptotic growth of
  the generating function. Even though our results are restricted to networks
  with a finite number of states, we show that the parabolic bound is also
  valid for three paradigmatic examples of driven diffusive systems for which
  the generating function can be calculated using the additivity principle.
  Our bounds provide a new general class of constraints for nonequilibrium
  systems.

\end{abstract}


\maketitle

\section{Introduction}

Equilibrium statistical physics is governed by a universal principle stating
that in an isolated system each microstate is equally likely.  For a system in
contact with a heat bath, thus the famous Gibbs-Boltzmann distribution arises
that involves only the Hamiltonian of the system and the temperature of the
bath. In non-equilibrium, a similarly universal principle is not known and may
not even exist. One characteristic feature of non-equilibrium systems is that
they necessarily come with dissipation, i.e., entropy production.
Non-equilibrium steady states, generated by time-independent driving have a
constant average entropy production. Observed for a finite time, the entropy
change exhibits fluctuations that are universally constrained by the
fluctuation theorem
\cite{evan93,evan94,gall95,kurc98,lebo99,seif12}, which
is arguably the most universal principle discovered for non-equilibrium
systems so far.

The fluctuation theorem relates the probability to observe a negative entropy
change to the one for observing the corresponding positive value. In this
sense, it constrains ``half'' of the distribution. Experiments have
illustrated and tested this symmetry, \textit{inter alia}, for colloidal particles
\cite{spec07,mehl12,gome11a}, energy exchange between two conductors \cite{cili13}, 
small electronic systems at low temperature \cite{peko15}, molecular motors \cite{haya10}, 
and shaken granular matter \cite{feit04,kuma11,joub12}.
In a refined version, the fluctuation theorem holds not only for entropy
change but also for the joint probability of all currents contributing to
the entropy change \cite{lebo99,andr07b}, which involves the
corresponding affinities like non-conservative forces for colloids, chemical
potential differences for bio-molecular reactions or voltage drops for
electronic circuits. Generally, these individual currents in a multi-cyclic network,
however, are not restricted by the fluctuation theorem or any other universal
result.

In this paper, we introduce a complementary class of constraints, not only on
the distribution of entropy change, but of any individual current in a
network. These constraints universally
bound the fluctuations over the full range of positive and negative values, in
particular the extreme fluctuations.  The crucial parameters
characterizing these bounds are the average entropy production, the affinities,
topological features like the number of states in a cycle and the activity,
i.e., the average number of transitions per unit time. 
If one knows such parameters, current fluctuations can be bounded
independently of the specific transition rates. Correspondingly, a measurement
of such current fluctuations will make it possible to infer constraints on these
parameters which in an experiment may not be known or not be directly
accessible.  This study substantially extends and generalizes work in which we
have recently explored universal relations between dissipation and 
dispersion of currents leading to a general thermodynamic uncertainty
relation \cite{bara15} and allowing the inference of topological properties of
enzymatic networks \cite{bara15a,bara15c}.

We employ the formalism of large deviations
\cite{ellis,touc09} in which for large times the exponential decay of the
tails of the distribution function is characterized by a rate function. This
rate function can be obtained from the Legendre transformation of the scaled
cumulant generating function.  For the latter, we derive 
a series of lower bounds that can be divided
into four classes: a parabolic bound, a hyperbolic cosine bound, an
exponential bound, and an asymptotic bound relevant for large values of $z$,
where $z$ is the real variable in the scaled cumulant generating function.
The last two bounds can be proved exactly while the first two are conjectures
based on extensive numerics. These universal bounds are valid for any
nonequilibrium system described by a Markov process with a finite number of
states. 

The hydrodynamic fluctuation theory for driven diffusive systems in contact with two
reservoirs by Bertini \textit{et al.} \cite{bert01a,bert02b,bert15} has been another
major development in nonequilibrium statistical physics. This theory leads to
a (typically hard) variational problem that, if solved, leads to the exact
rate function of the current of particles or heat between reservoirs. The
additivity principle derived in \cite{bodi04} is a more direct method that
allows for the calculation of the scaled cumulant generating function related
to the current in driven diffusive systems. For
example, this method has been used to calculate this function for the
symmetric simple exclusion process (SSEP) \cite{bodi04,derr07}, the
Kipnis-Marchioro-Pressuti (KMP) model \cite{hurt09,hurt10}, and the weakly
asymmetric simple exclusion process (WASEP) \cite{gori12}.  These results are
valid in the limit of large system size, for which the number of states
diverges.  Even though our bounds are restricted to the case of a finite
number of states, we show that the scaled cumulant generating functions
obtained from the additivity principle for these three models lies inside our
parabolic bound.

The paper is divided as follows. In Sec.~\ref{sec:definitions} we define the
entropy, the currents and their generating functions. Our main results are
summarized in Sec.~\ref{sec:sum}. Secs.~\ref{sec:parabolic}, \ref{sec:cosh}, \ref{sec:exponential},
and \ref{sec:asymptotics} contain the parabolic, hyperbolic cosine, exponential, and
asymptotic bounds, respectively.  In Sec.~\ref{sec:1D}, the parabolic bound is
compared with exact results for the SSEP, the KMP model, and the WASEP. We
conclude in Sec.~\ref{sec:conclusion}.  The \hyperref[sec:expproof]{appendices}
contain various proofs and details on the numerics.

\section{Large deviations in Markovian networks}
\label{sec:definitions}

We consider a Markovian network consisting of $N$ discrete states $\{i\}$ and
allow for transitions with rates $k_{ij}\geq 0$ from state $i$ to $j$. All
transitions are taken to be reversible, i.e., $k_{ij}>0$ implies
$k_{ji}>0$. The time dependent probability distribution $p_i(t)$ of state $i$
at time $t$ evolves according to the master equation
\begin{equation}
  \partial_t p_i(t)=\sum_j \mathcal{L}_{ij} p_j(t)\equiv\sum_j[k_{ji}-r_i\delta_{ij}]p_j(t),
\end{equation}
where the exit rate from state $i$ is defined as
\begin{equation}
  r_i\equiv\sum_\ell k_{i\ell}.
\end{equation}
For large times $t$, $p_i(t)$ tends to the stationary distribution $\pstat_i$,
which satisfies $\sum_j \mathcal L_{ij}\pstat_i=0$.

\floatsetup[figure]{style=plain,subcapbesideposition=top}
\begin{figure} 
  \centering
  \includegraphics{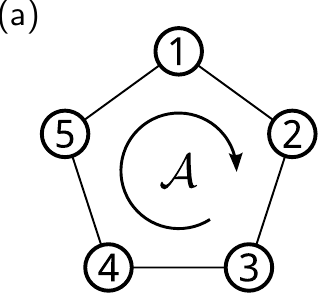}
  \includegraphics{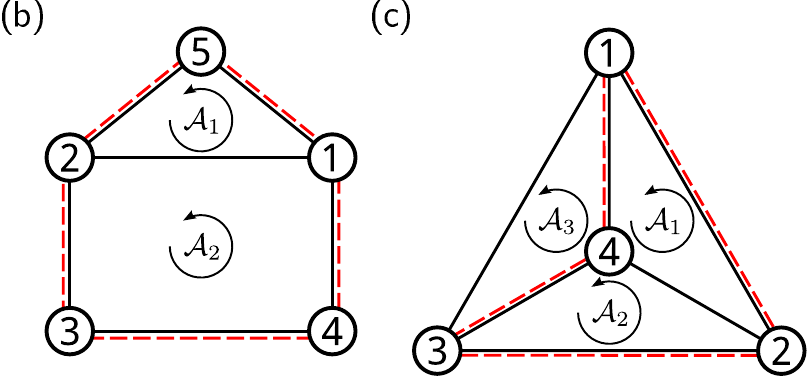}
  \caption{(a) Unicyclic network with affinity $\mathcal{A}$ and five
    states. (b) Multicyclic network with two fundamental cycles, one with
    three states and affinity $\mathcal{A}_1$ and the other with four states
    and affinity $\mathcal{A}_2$.  The red dashed lines indicate a
    cycle with affinity $\mathcal{A}_1+\mathcal{A}_2$ and five states. (c) Multicyclic
    network with three fundamental cycles with three states each. The affinities of these cycles are $\mathcal{A}_1$, $\mathcal{A}_2$, and
    $\mathcal{A}_3$. The red dashed lines indicate a
    cycle with affinity $\mathcal{A}_1+\mathcal{A}_2$ and four
    states.}
  \label{fig:networks}
\end{figure}
Following Schnakenberg, we identify a complete set of fundamental cycles
$\{\beta\}$ within the network \cite{schn76}. Each cycle is associated with an affinity
$\mathcal{A}_{\beta}$, a fluctuating current $X_\beta(t)$ that counts cycle
completions after time $t$ (the so-called \textit{integrated current}) and an
average current 
\begin{equation}
 J_\beta\equiv\mean{X_\beta(t)}/t,
\end{equation}
where the brackets indicate an average over stochastic trajectories.
The average is independent of $t$ for initial conditions drawn from the steady
state distribution.  

Upon a transition $i\to j$, $X_\beta$ increases by the
generalized distance $d_{ij}^\beta=-d_{ji}^\beta$. These increments are
constrained to add up to one for every closed loop that completes the cycle
once in forward direction. For example, $X_\beta$ could be the scaled displacement of
a molecular motor. The affinity $\mathcal{A}_\beta$ would then be given by the
external force times the length of a full motor step, while $d_{ij}^\beta$
denotes the relative length of a sub-step related to the conformational change
$i\to j$. The ratio of forward and backward transition rates fulfill the local
detailed balance relation
\begin{equation}
  \ln(k_{ij}/k_{ji})=\sum_\beta d_{ij}^\beta\mathcal{A}_\beta+E_i-E_j,
\end{equation}
where $E_i$ denotes the equilibrium free energy associated with state $i$, 
and, in general, the sum over $\beta$ is a sum over all fundamental cycles in the network of
states \cite{schn76} (see also \cite{andr07b,bara12a} for a precise definition of a fundamental cycle). For
notational convenience we have set Boltzmann's constant $k_\mathrm{B}$ and the
temperature $T$ to unity, thus energies and affinities are given in units of
the thermal energy $k_\mathrm{B} T$. The (fluctuating) entropy change in the surrounding
medium $\sm(t)$ is given by the increments
$d_{ij}^\mathrm{s}=\ln(k_{ij}/k_{ji})$. The average entropy production reads
\begin{equation}
 \mep\equiv  \mean{\sm(t)}/t=\sum_\beta\mathcal{A}_\beta J_\beta.
\label{eq:medium_entropy}
\end{equation}

Adopting a vector notation $\vec X$ for the set of all cycle currents
$X_\beta$ the scaled cumulant generating function is defined as
\begin{equation}
   \lambda(\vec z)\equiv\lim_{t\to\infty}\frac{1}{t}\ln\mean{\exp[\vec
     z\cdot\vec X(t)]},
   \label{eq:gen-func-md}
\end{equation}
where $\vec z$ is a real vector.
As an abbreviation we will refer to $\lambda(\vec z)$ simply as the ``generating function''.
It can be shown that $\lambda(\vec z)$ is the largest eigenvalue 
of the modified Markov generator $\mathcal{L}_{ij}(\vec z)$, which is defined
as \cite{lebo99,koza99}
\begin{equation}
   \mathcal{L}_{ij}(\vec z)\equiv
   \mathcal{L}_{ij}\exp(\vec{z}\cdot\vec{d}_{ji}),
   \label{eq:defL}
\end{equation}
where $\vec{d}_{ji}$ is a vector with components $d_{ji}^\beta$.
The variable $X_\beta/t$, or more conveniently the scaled variable
\begin{equation}
  \xi_\beta\equiv X_\beta/(tJ_\beta),
  \label{eq:defxi}
\end{equation}
satisfies a large deviation principle \cite{touc09,chet13} of the form
\begin{equation}
  \mathrm{Prob}(\vec X,t)\sim\exp[-th(\vec\xi)]
\end{equation}
with a rate function $h(\vec \xi)$ that is given by the Legendre-Fenchel transform
\begin{equation}
  h(\vec\xi)=\max_{\vec z}\left[\sum_\beta z_\beta
  J_\beta\xi_\beta-\lambda(\vec z)\right].
\end{equation}

The fluctuation theorem is a symmetry, known as Gallavotti-Cohen symmetry, on the generating 
function of the form \cite{lebo99,andr07b}
\begin{equation}
  \lambda(\vec z)=\lambda(-\vec{\mathcal{A}}-\vec{z}).
  \label{eq:GC_sym}
\end{equation}
In terms of the rate function this symmetry reads 
\begin{equation}
  -h(\vec{\xi})+h(-\vec{\xi})=\sum_\beta \xi_\beta J_\beta\mathcal{A}_\beta.
\end{equation}
The generating function related to a single fluctuating current $X_\alpha$ is
\begin{equation}
  \lambda_\alpha(z)\equiv\lambda(z\vec{e}_\alpha)=\lim_{t\to\infty}\frac{1}{t}\ln\mean{\exp[zX_\alpha(t)]},
\label{eq:gen-func-cycle}
\end{equation}
where $\vec{e}_\alpha$ is the unit vector associated with the current in cycle
$\alpha$.  Generally, this function does not exhibit a symmetry of the form
\eqref{eq:GC_sym} as extensively discussed in \cite{bara12a} (see also \cite{bara10,bara11}).
In contrast, the evaluation of $\lambda(\vec z)$ along the vector
$\vec{\mathcal{A}}$ yields
\begin{equation}
  \lambda_s(z)\equiv\lambda(z\vec{\mathcal{A}})=\lim_{t\to\infty}\frac{1}{t}\ln\mean{\exp[zs_\mathrm{m}(t)]}
\end{equation}
as the generating function of the entropy change. It is symmetric with
respect to $z=-1/2$, which expresses the fluctuation theorem that holds for
this observable.
The rate functions associated with the probability distributions of these variables read
\begin{equation}
  h_\alpha(\xi_\alpha)=\max_z[z J_\alpha\xi_\alpha-\lambda_\alpha(z)]
\end{equation}
and, introducing the scaled entropy change $s\equiv \sm(t)/(\mep t)$
in analogy to Eq.~\eqref{eq:defxi},
\begin{equation}
  h_s(s)=\max_z[z \mep s-\lambda_s(z)],
  \label{eq:rf_hs}
\end{equation}
respectively. 

An important distinction in this paper is the one between unicyclic and
multicyclic networks of states, illustrated in Fig.~\ref{fig:networks}.  For
unicyclic networks, where there is only a single affinity
$\mathcal{A}\equiv\mathcal{A}_\alpha$ and a single fluctuating current
$X\equiv X_\alpha$, we no longer have to distinguish between the different
types of generating functions and can simply write
\begin{equation}
  \lambda(z)\equiv\lambda_\alpha(z)=\lambda_s(z/\mathcal{A}).
\end{equation}

In the following, we will be interested in functions $b(\vec z)$ that bound the
generating function $\lambda(\vec z)$ from below, i.e.,
\begin{equation}
  b(\vec z)\leq \lambda(\vec z)
  \label{eq:bound-general}
\end{equation}
for all $\vec{z}$.  As special cases, the relation \eqref{eq:bound-general}
can be used to extract bounds for individual fluctuating currents, $\lambda_\alpha(z)\geq
b_\alpha(z)\equiv b(z\vec{e}_\alpha)$, and the entropy change,
$\lambda_s(z)\geq b_s(z)\equiv b(z\vec{\mathcal{A}})$. Such bounds immediately
imply upper bounds on the rate functions
\begin{equation}
  h_\alpha(\xi_\alpha)\leq\max_{\vec{z}}[z J_\alpha\xi_\alpha-b_\alpha(z)].
  \label{eq:ldf_bound}
\end{equation}
and
\begin{equation}
  h_s(s)\leq\max_{\vec{z}}[z \mep s-b_s(z)].
  \label{eq:ldf_bound_s}
\end{equation}

For any generating function the coefficients of the Taylor
expansion around  $z=0$ correspond to the cumulants. The Fano factor that quantifies the dispersion of the distribution is defined as 
\begin{equation}
  F\equiv
  \lim_{t\to\infty}\frac{\mean{X(t)^2}-\mean{X(t)}^2}{\mean{X(t)}}=\frac{\lambda''(0)}{\lambda'(0)},
\end{equation}
where $X$ is a random variable. We denote the Fano factor associated with an individual current by $F_\alpha$ and the one 
associated with the entropy change in the medium by $F_s$. Since global lower bounds $b(z)$ with $b(0)=0$ must share a tangent with $\lambda(z)$ at $z=0$ while having a
stronger curvature, every such bound implies with
\begin{equation}
  F\geq\frac{b''(0)}{b'(0)}
\label{eq:FFbound}
\end{equation}
a bound on the Fano factor.

Since $\lambda(0)=0$ holds trivially for all networks, we usually require that
our bounds are saturated for $\vec{z}=0$. This requirement will only be
lifted for a bound that captures the asymptotic behavior for large $|\vec z|$ in
Sec.~\ref{sec:asymptotics}. Hence, if $\lambda(\vec z)$ is analytic, $b(\vec
z)$ must have the same gradient as $\lambda(\vec z)$.

\section{Summary of main results}
\label{sec:sum}

\begin{figure} 
  \centering
  \includegraphics{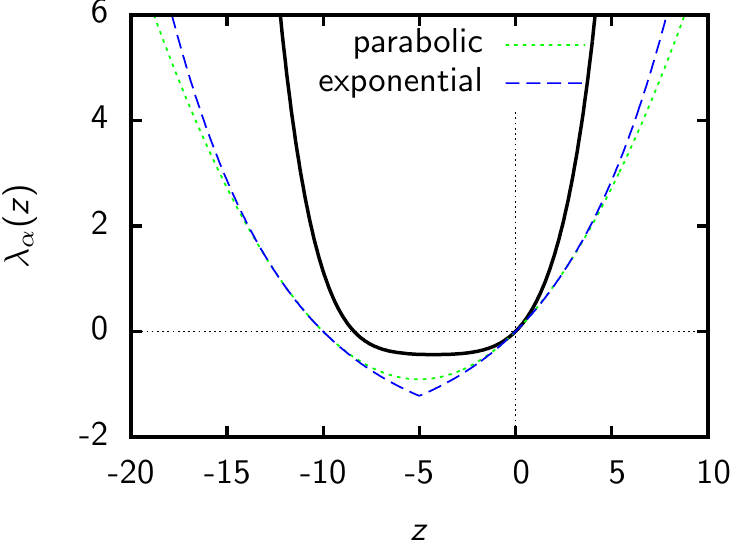}
  \caption{Illustration of our two main results for the generating function of
  an individual cycle current (thick black curve). The parabolic bound that depends on 
  the entropy production is shown as a green dotted line and
  the exponential bound that depends on the entropy production and the
  activity is shown as a blue dashed line. 
  The generating function refers to the
  cycle $\alpha=1$ of the ``house-shaped'' network shown in
  Fig.~\ref{fig:networks}b. The affinities are $\mathcal{A}_1=8$ and
  $\mathcal{A}_2=6$, all transition rates were set to 1 except for
  $k_{15}=\exp(\mathcal{A}_1/2)$, $k_{51}=\exp(-\mathcal{A}_1/2)$,
  $k_{41}=\exp(\mathcal{A}_2/2)$, and $k_{14}=\exp(-\mathcal{A}_2/2)$. The
  quantities relevant for the bounds are the average current $J_\alpha\simeq0.36$,
  the entropy production $\sigma\simeq 3.6$, and the activity $R\simeq2.13$.}
  \label{fig:mainresults}
\end{figure}

The two main bounds on the generating function obtained in this paper are
illustrated in Fig.~\ref{fig:mainresults}. First, the generating function
$\lambda(\vec z)$ for any network is bounded by a parabola according to
\begin{equation}
    \lambda(\vec z)\geq \vec z\cdot \vec J\,(1+\vec z\cdot \vec J /\mep).
\end{equation}
This parabolic bound depends only on the average entropy production.  Second,
$\lambda(\vec z)$ is also bounded from below by an exponential function of the form
\begin{equation}
    \lambda(\vec z)\geq R\left[\mre^{(|\mep/2+\vec z\cdot\vec
    J|-\mep/2)/R}-1\right].
\end{equation}
This second bound depends on the average entropy production and on the
activity $R\equiv\sum_i\pstat_ir_i$, which is the average number of
transitions per time in the whole network.

Choosing a specific direction for the vector $\vec z$, both bounds are valid both
for any individual fluctuating current and for the entropy change. It is quite
remarkable that the fluctuations of any current in an arbitrary
multicyclic network can be bounded by a function involving only the average
entropy production in the case of the parabolic bound and the average entropy
production and activity in the case of the exponential bound. 
Even though there is no obvious relation between the parabolic bound and the
exponential bound, typically, the exponential bound becomes tighter than the
parabolic bound both for far from equilibrium conditions and for large $|\vec z|$.

We derive two further relevant lower bounds on $\lambda(\vec z)$ in this
paper. (1) A hyperbolic cosine bound, which is an extension of the parabolic
bound that is tighter and requires further knowledge of the affinities and the
topology of the network. (2) An asymptotic bound that becomes tight for large
values of $|\vec z|$ and requires knowledge of all transition rates.

These bounds
are complementary to the fluctuation theorem. Whereas they establish the
minimal value that $\lambda(\vec z)$ can take, the fluctuation theorem
constrains $\lambda(\vec z)$ to have the symmetry \eqref{eq:GC_sym}.

\section{Parabolic bound}
\label{sec:parabolic}
\subsection{Linear response regime}
In the limit of small affinities $\mathcal{A}_\beta$, the average current $J_\alpha$
depends linearly on the affinities,
\begin{equation}
  J_\alpha=\sum_\beta L_{\alpha\beta} \mathcal{A}_\beta
  \label{eq:onsagercurrent}
\end{equation}
with the symmetric and positive definite Onsager matrix
$L_{\alpha\beta}\equiv\partial
J_\alpha/\partial\mathcal{A}_\beta|_{\vec{\mathcal A}=0}$.  In the region
$\vec z\lesssim\ord{\vec{\mathcal{A}}}$, the generating function $\lambda(\vec
z)$ can be expanded as a quadratic form around its center of symmetry, which
is, due to Eq.~\eqref{eq:GC_sym}, located at $\vec
z=-\vec{\mathcal{A}}/2$. The requirement $\lambda(0)=0$ and
$\nabla\lambda(0)=\vec J$ fixes this expansion to
\begin{equation}
  \lambda(\vec
  z)=\sum_{\beta,\gamma}(z_\beta+{\mathcal{A}_\beta}/2)\,L_{\beta\gamma}\,(z_\gamma+{\mathcal{A}}_\gamma/2)-\mep/4,
\end{equation}
where the entropy production $\mep$ is given in Eq.~\eqref{eq:medium_entropy}.
Evaluating this function for $\vec{z}=z\vec{e}_\alpha$ yields as generating
function related to the individual current
\begin{equation}
  \lambda_\alpha(z)=zJ_\alpha+z^2 L_{\alpha\alpha}.
\end{equation}
The positive definiteness of the matrix
$G_{\beta\gamma}\equiv(L_{\alpha\alpha}L_{\beta\gamma}-L_{\alpha\beta}L_{\alpha\gamma})$
\cite{bara15suppl} (with $\alpha$ fixed) yields
\begin{equation}
  \sum_{\beta,\gamma}G_{\beta\gamma}\mathcal{A}_\beta\mathcal{A}_\gamma=L_{\alpha\alpha}\mep-J_\alpha^2\geq 0.
\end{equation}
Hence $\lambda_\alpha(z)$ is bounded from below by
\begin{equation}
 \lambda_\alpha(z)\geq z J_\alpha(1+z J_\alpha /\mep).  
\label{eq:lr_bound_lambda_alpha}
\end{equation}
Using the Legendre transform \eqref{eq:ldf_bound}, this bound can be
transformed into a bound for the rate function
\begin{equation}
  h_\alpha(\xi_\alpha)=\frac{L_{\alpha\alpha}}{4J_\alpha^2}(\xi_\alpha-1)^2\leq\frac{\mep}{4}(\xi_\alpha-1)^2.
  \label{eq:lr_bound_h_alpha}
\end{equation}
Since the direction $\vec{e}_\alpha$ can be chosen arbitrarily, the bound
\eqref{eq:lr_bound_lambda_alpha} can be stated in a multidimensional
formulation as
\begin{equation}
  \lambda(\vec z)\geq \vec z\cdot \vec J\,(1+\vec z\cdot \vec J /\mep). 
 \label{eq:lr_bound_lambda}
\end{equation}
Equality holds along the line $\vec z\propto\vec{\mathcal{A}}$, which
corresponds to the generating function $\lambda_s(z)=\lambda(\mathcal{A} z)$
associated with entropy change. Within linear response, the rate function for
the scaled entropy change $s$ is thus given by
\begin{equation}
  h_s(s)=\frac{\mep}{4}(s-1)^2.
\end{equation}

Eq.~\eqref{eq:lr_bound_h_alpha} shows that the knowledge of the average entropy production is sufficient to bound
the whole range of fluctuations of any individual current in the linear response regime. Surprisingly, as we show 
next, this parabolic bound is also valid beyond the linear response regime. 

The parabolic bound has also an important
  consequence for fluctuations in systems at equilibrium. To study this case it is more
  convenient to scale the fluctuating currents as $x_\beta\equiv X_\beta/t
  =J_\beta\xi_\beta$. For the corresponding rate function $\tilde
  h_\alpha(x_\alpha)=h_\alpha(x_\alpha J_\alpha)$, the bound
  \eqref{eq:lr_bound_h_alpha} then reads 
  \begin{equation}
       \tilde
       h_\alpha(x_\alpha)\leq\frac{1}{4}\frac{\sum_{\gamma\delta}L_{\gamma\delta}\mathcal{A}_\gamma\mathcal{A}_\delta}{\left(\sum_\beta
        L_{\alpha\beta}\mathcal{A}_\beta\right)^2}x_\alpha^2+\mathcal{O}(\vec{\mathcal{A}}),
    \label{eq:parabolic_eq}
  \end{equation}
for small $\mathcal{A}_\beta$ with fixed $x_\alpha$.
Here, we have represented the average currents using
Eq.~\eqref{eq:onsagercurrent}. This bound is supported by our numerics
presented in appendix~\ref{sec:numerics} as, where we have checked \eqref{eq:lr_bound_lambda_alpha} also for
$\vec{z}\gtrsim\mathcal{O}(\vec{\mathcal{A}})$. For multicyclic
networks at equilibrium, the prefactor in \eqref{eq:parabolic_eq} depends on
the direction in which the limit $\vec{\mathcal{A}}\to 0$ is
taken. In particular, choosing $\vec{\mathcal{A}}\propto \vec{e}_\alpha$ yields
\begin{equation}
  \tilde h_\alpha(x_\alpha)\leq  x_\alpha^2/(4 D_\alpha).
  \label{eq:parabolic_eq_D}
\end{equation}
Thus, the equilibrium fluctuations of any current $X_\alpha$ can be
bounded by the parabola that is defined as the continuation of the
quadratic expansion of the rate function around $x_\alpha=0$. In other
words, the Gaussian approximation for typical fluctuations always
underestimates the probability of extreme fluctuations in equilibrium
systems. Since this bound is exact for small $x_\alpha$, performing
the limit $\vec{\mathcal{A}}\to 0$ in a direction different from
$\vec{e}_\alpha$ cannot yield a stronger bound.

\subsection{Beyond linear response: Unicyclic case}
The parabolic shape of the generating function for
  $z\lesssim\mathcal{O}(\mathcal{A})$ and of the rate function for
  $\xi\lesssim\mathcal{O}(1)$ can be regarded as a signature of linear
response. It arises only for nearly vanishing affinities or for freely
diffusing particles, where the linearity between affinity and current persists
even for high affinities. Beyond this regime, one universally observes two
characteristic changes in the rate function
\cite{mehl08,doro11,spec12}. First, the tails for large values of
$|\xi_\alpha|$ grow no longer quadratically but with a scaling somewhere
between linear and quadratic. Second, there is a formation of a ``kink''
around the value $\xi_\alpha=0$. For finite numbers of states, the rate
function is still analytic in this region, but it exhibits a significantly
enhanced curvature.  In the Legendre transformed picture of the generating
function $\lambda_\alpha(z)$, these two effects show up as a faster than
quadratic growth for large $z$ and a pronounced plateau around the minimum of
$\lambda_\alpha(z)$.

\begin{figure} 
  \centering
  \includegraphics{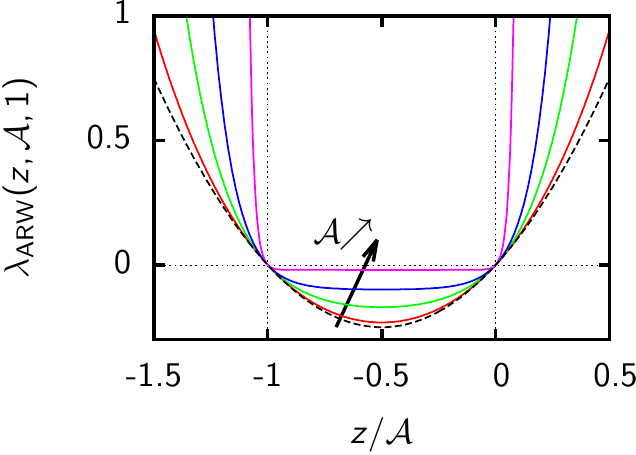}
  \includegraphics{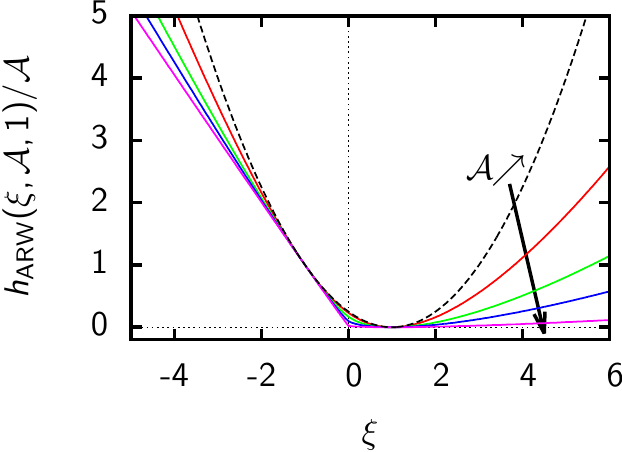}
  \caption{The generating function $\lambda(z)$ and the rate function
    $h(\xi)$ of the asymmetric random walk for selected affinities
    $\mathcal{A}$ ($2,5,10,50$) and $N=1$. Black arrows indicate the
    direction of increasing $\mathcal{A}$. The parabolic bound for the generating function 
    \eqref{eq:bound_parabolic_uc} and for the rate function \eqref{eq:bound_parabolic_uc_ldf} are shown as
    black dashed curves.}
  \label{fig:ldf_arw}
\end{figure}
This behavior of the generating function is best illustrated with an asymmetric random walk (ARW), as shown in Fig.~\ref{fig:ldf_arw}. Consider
a network consisting of a single cycle with $N$ vertices and affinity
$\mathcal{A}$, as shown in Fig.~\ref{fig:networks}a. The hopping rates in
forward and backward directions $k^+$ and $k^-$ are uniform with
\begin{equation}
  \ln\frac{k^+}{k^-}=\mathcal{A}/N.
\end{equation}
The average current in this model is $J=(k^+-k^-)/N$ and the entropy production is $\mep=J\mathcal{A}$.
It can be shown that the generating function is given by \cite{lebo99}
\begin{align}
  \lambda(z)&=k^+\left[\mre^{z/N}+\mre^{-(z+\mathcal{A})/N}-1-\mre^{-\mathcal{A}/N}\right]
  \label{eq:gen_func_arw}\\
&=J \lambda_\mathrm{ARW}(z,\mathcal{A},N), \nonumber
\end{align}
where
\begin{equation}
\lambda_\mathrm{ARW}(z,\mathcal{A},N)\equiv
\frac{\cosh[(z+\mathcal{A}/2)/N]-\cosh[\mathcal{A}/(2N)]}{(1/N)\sinh[\mathcal{A}/(2N)]}.
\label{eq:arwdef}
\end{equation}
Similarly, the rate function corresponding to the generating function
\eqref{eq:gen_func_arw} is given by \cite{lebo99}
\begin{equation}
  h(\xi)=J\,h_\mathrm{ARW}(\xi,\mathcal{A},N),
\end{equation}
where
\begin{align}
h_\mathrm{ARW}(\xi,\mathcal{A},N)\equiv\frac{N}{\sinh[\mathcal{A}/(2N)]}\Big[a\xi\operatorname{arsinh}(a\xi)-a\xi\frac{\mathcal{A}}{2N}\nonumber\\
-\sqrt{1+(a\xi)^2}+\sqrt{1+a^2}\Big]
\label{eq:ldf_arw}  
\end{align}
and $a\equiv\sinh[\mathcal{A}/(2N)]$. 
As shown in Fig.~\ref{fig:ldf_arw}, the generating function \eqref{eq:gen_func_arw} is bounded from below by the parabola
\begin{equation}
  \lambda_{\mathrm{ARW}}(z,\mathcal{A},N)\geq z\,J\,(1+z/\mathcal{A}),
  \label{eq:bound_parabolic_uc}
\end{equation}
and the rate function is bounded from above by the parabola 
\begin{equation}
  h_{\mathrm{ARW}}(z,\mathcal{A},N)\leq \mathcal{A}(\xi-1)^2/4.
  \label{eq:bound_parabolic_uc_ldf}
\end{equation}

In Sec.~\ref{sec:cosh}, we will show in the context of an even stronger,
affinity-dependent bound, that the bound \eqref{eq:bound_parabolic_uc} holds
also for arbitrary unicyclic networks with non-uniform transition rates.

\subsection{Beyond linear response: multicyclic case}

Based on numerical evidence we conjecture that 
\begin{equation}
    \lambda(\vec z)\geq \vec z\cdot \vec J\,(1+\vec z\cdot \vec J /\mep)
   \label{eq:gf_bound_parabolic}
\end{equation}
holds globally for all vectors $\vec z$ and for all types of Markovian networks. 
In terms of the individual current in a cycle $\alpha$, this conjecture can be
formulated as
\begin{equation}
\lambda_\alpha(z)\geq z J_\alpha(1+z J_\alpha /\mep)
   \label{eq:gf_bound_parabolic_ic}
\end{equation}
whereas for the entropy change
\begin{equation}
  \lambda_s(z)\geq z\mep(1+z).
   \label{eq:gf_bound_parabolic_entropy}
\end{equation}
The bound \eqref{eq:gf_bound_parabolic} becomes the same as
\eqref{eq:lr_bound_lambda} in the linear response regime. However,  \eqref{eq:gf_bound_parabolic} is also valid beyond this regime where  the currents $\vec{J}$  are the actual
average currents in the steady state, as determined from $\nabla \lambda(0)$,
which are different from the linear response currents \eqref{eq:onsagercurrent}.  

\begin{figure} 
  \centering
  \includegraphics{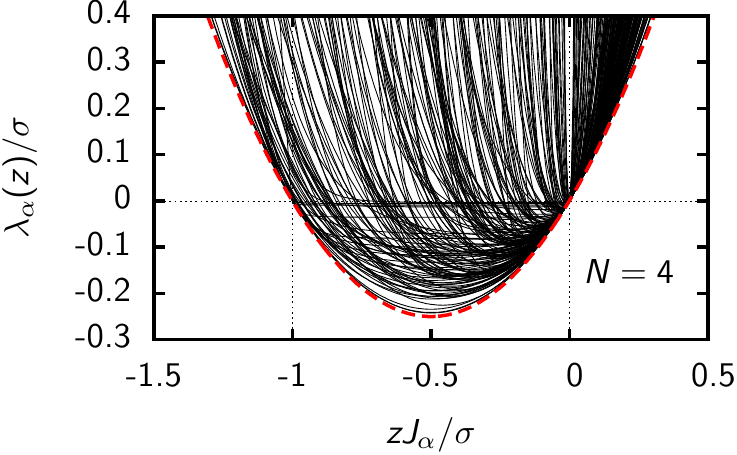}
  \includegraphics{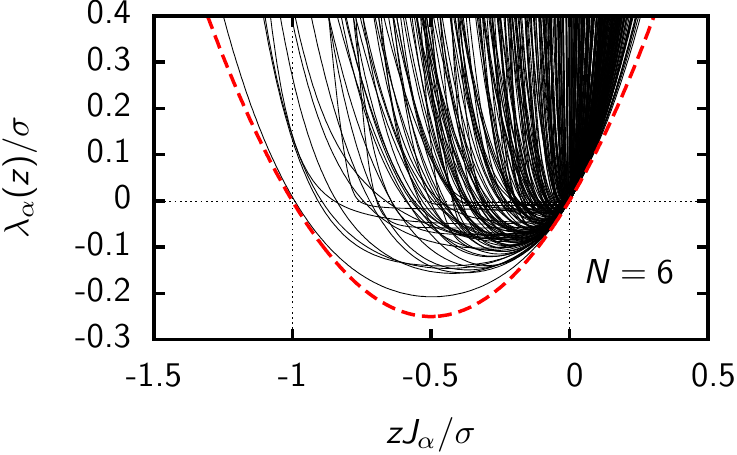}
  \caption{Generating functions $\lambda_\alpha(z)$ for an individual current
    in fully connected networks with random transition rates. The
      black curves in the upper and lower panels
    correspond to networks with $N=4$ and $N=6$ vertices, respectively. The
    parabolic bound \eqref{eq:gf_bound_parabolic_ic} is shown as a dashed curve.}
  \label{fig:bound_parabolic}
\end{figure}

The numerical evidence for this bound is illustrated in
Fig.~\ref{fig:bound_parabolic}.  We generated a large set of networks of
states with random transition rates, drawn according to the procedure
described in appendix~\ref{sec:numerics}.  As the affinity increases, generating functions
globally deviate in a positive direction from the parabolic shape. Only at the
trivial points $\vec z=0$ and $\vec z=-\vec{\mathcal A}$ does the generating
function in Eq.~\eqref{eq:gf_bound_parabolic} acquire with zero the same
value for all networks.  For
larger networks ($N=6$) the left hand side of the plot becomes less populated,
since the probability of the vectors $\vec{e}_\alpha$ and $\vec{\mathcal{A}}$
being nearly parallel becomes smaller in higher dimensions.  The full
numerical evidence for this parabolic bound is explained in appendix~\ref{sec:numerics}.

The local evaluation \eqref{eq:FFbound} of the parabolic bound \eqref{eq:gf_bound_parabolic_ic} for an individual current yields the relation
\begin{equation}
F_\alpha\mep/J_\alpha\geq 2
 \label{eq:uncthermo} 
\end{equation}
for the Fano factor associated with the current $X_\alpha$.
This ``thermodynamic uncertainty relation'', which imposes a minimal energetic
cost that must be paid for small uncertainty in the output of an enzymatic
reaction, has been derived in \cite{bara15}. Hence, the parabolic bound is a
generalization of this relation.  From relation \eqref{eq:uncthermo},
measurements of the dispersion and average of an individual current can
provide a lower bound on the average entropy production $\sigma\ge
2J_\alpha/F_\alpha$. This bound makes it possible to estimate the entropy
production by measuring a single individual current
\cite{rold10,rold12}. Applied to the entropy production in
the medium, the
parabolic bound \eqref{eq:gf_bound_parabolic_entropy} leads to
\begin{equation}
  F_s\geq 2.
\end{equation}

\section{Hyperbolic cosine bound}
\label{sec:cosh}

\subsection{Unicyclic networks}

For a unicyclic network, the parabolic bound is saturated in the linear response regime.
As shown in Fig.~\ref{fig:ldf_arw} for the asymmetric random walk, the generating function deviates more from the parabolic bound
as the affinity $\mathcal{A}$ increases. We now 
discuss an affinity dependent bound that is stronger than the parabolic bound. This affinity 
dependent bound is also less universal as it requires the knowledge of
$\mathcal{A}$. For example, in a biochemical network a fixed affinity means that the chemical potential difference driving a chemical reaction is
known.

\begin{figure} 
  \centering
  \includegraphics{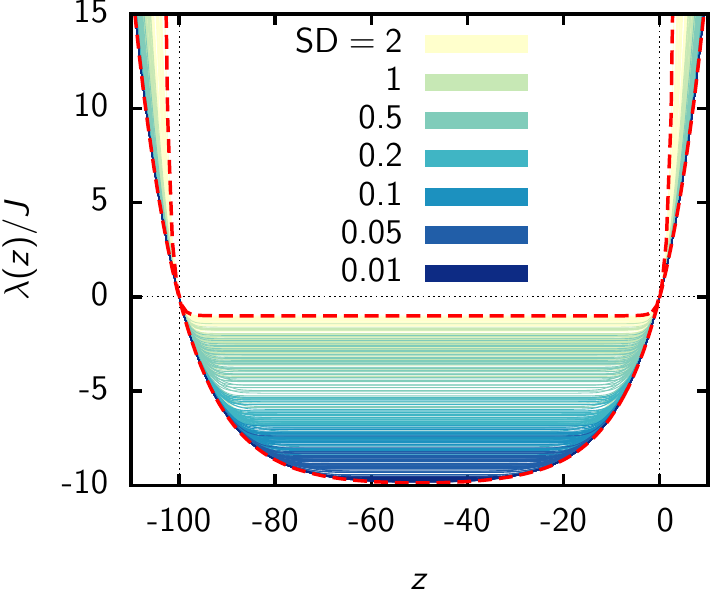}
  \caption{Generating function $\lambda(z)$ scaled by the steady state current
    $J$ in unicyclic networks with $N=10$ states and fixed affinity
    $\mathcal{A}=100$. The gray-scale (color-scale) encodes the standard
    deviation (SD) used for sampling the transition rates (see
    appendix~\ref{sec:numerics}). Dark gray (blue) corresponds to a
    nearly uniform distribution of transition rates and light gray
      (yellow) to a broad distribution of transition rates. The
      lower and upper bounds \eqref{eq:cosh-bound-lower} and
      \eqref{eq:cosh-bound-upper}, respectively, are shown as (red) dashed lines.}
  \label{fig:bounds_uc_highA}
\end{figure}

The transition rates for an arbitrary unicyclic model with $N$ states and periodic boundary conditions are denoted by  
\begin{equation}
k_{i,i+1}=k^+_i\mathrm{\ and\ }k_{i,i-1}=k^-_i,
\label{eq:unif_rates}
\end{equation}
where $i=1,2,\ldots,N$. A fixed affinity $\mathcal{A}$ implies the constraint
\begin{equation}
\frac{\prod_{i=1}^N k_i^+}{\prod_{i=1}^N k_i^-}=\mre^\mathcal{A}
\label{eq:uniconst}
\end{equation}
on the transition rates.
Different choices of the transition rates that fulfill this restriction can lead to different generating functions,
as shown in Fig.~\ref{fig:bounds_uc_highA}. In particular, if the transition rates are uniform, i.e., $k_{i}^+=k^+$ 
and $k_{i}^-=k^-$ the generating function divided by the average current $\lambda(z)/J$ becomes  $\lambda_\mathrm{ARW}(z,\mathcal{A},N)$,
which is given in Eq.~\eqref{eq:arwdef}.  The opposite extreme choice for the transition rates is the case where the network behaves effectively like
there was only one link between states ($N=1$) and all the affinity is concentrated in this
single link. In this case $\lambda(z)/J$ becomes $\lambda_\mathrm{ARW}(z,\mathcal{A},1)$, which fulfills $\lambda_\mathrm{ARW}(z,\mathcal{A},1)\geq\lambda_\mathrm{ARW}(z,\mathcal{A},N)$. 

From these considerations we conjecture that for unicyclic networks 
\begin{equation}
 \lambda_\mathrm{ARW}(z,\mathcal{A},1)\geq\lambda(z)/J\geq\lambda_\mathrm{ARW}(z,\mathcal{A},N).
 \label{eq:inecosh}
\end{equation}
Hence, using the definition \eqref{eq:arwdef}, Eq.~\eqref{eq:inecosh} implies the lower bound
\begin{equation}
  \label{eq:cosh-bound-lower}
  \lambda(z)\geq J\,\frac{\cosh[(z+\mathcal{A}/2)/N]-\cosh[\mathcal{A}/(2N)]}{(1/N)\sinh[\mathcal{A}/(2N)]},
\end{equation}
which we call the hyperbolic cosine bound for a unicyclic network. This conjectured bound  
is supported by the numerical evidence shown in Fig.~\ref{fig:bounds_uc_highA}. Eq. \eqref{eq:inecosh} also leads to an upper bound
\begin{equation}
  \label{eq:cosh-bound-upper}
  \lambda(z)\leq J\,\frac{\cosh(z+\mathcal{A}/2)-\cosh(\mathcal{A}/2)}{\sinh(\mathcal{A}/2)}.
\end{equation}
A rigorous proof for this upper bound is provided in appendix~\ref{sec:upperbound}.

In Fig.~\ref{fig:bounds_uc}, we show the bounds \eqref{eq:cosh-bound-lower}
and \eqref{eq:cosh-bound-upper} for different values of the affinity. For smaller $\mathcal{A}$
the bounds are closer to each other. In the linear response regime, up to quadratic order in $z$, they become the same and equal to the parabolic bound
in Eq.~\eqref{eq:bound_parabolic_uc}.

\begin{figure} 
  \centering
  \includegraphics{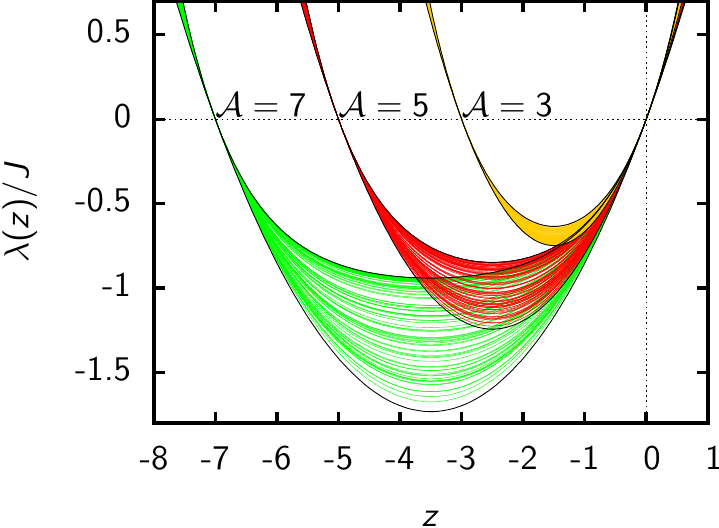}
  \caption{Generating function $\lambda(z)$ scaled by the steady state
    current $J$ in unicyclic networks with $N=10$ for three families of distinct
    affinities $\mathcal{A}$. For each family, transition rates were sampled
    according to the procedure described in appendix~\ref{sec:numerics}. The black
    curves refer to the lower and upper bound from
    Eqs.~\eqref{eq:cosh-bound-lower} and \eqref{eq:cosh-bound-upper}.}
  \label{fig:bounds_uc}
\end{figure}

The bounds in Eqs.~\eqref{eq:cosh-bound-lower} and \eqref{eq:cosh-bound-upper}
lead to the bounds
\begin{equation}
\coth\left(\frac{\mathcal{A}}{2}\right)\ge F \geq \frac{1}{N}\coth\left(\frac{\mathcal{A}}{2N}\right)
\label{eq:FF_bound_coth2}
\end{equation}
on the Fano factor defined in Eq.~\eqref{eq:FFbound}.
The lower bound is an affinity dependent bound on the Fano factor that has been obtained in \cite{bara15}. In the formal limit  $\mathcal{A}\to \infty$
it becomes $F\ge 1/N$. This bound for formally divergent affinity is a key
result in statistical kinetics \cite{moff14} as it allows for an estimate
on the number of states $N$ from measurements of the Fano factor. The upper bound on $F$ in Eq.~\eqref{eq:FF_bound_coth2} is a new result. 
It is a generalization of the known result $F\leq 1$, which is  also valid in the limit $\mathcal{A}\to \infty$ \cite{moff14}.

For systems at equilibrium, the bounds \eqref{eq:inecosh} become
\begin{equation}
  2D[\cosh(z)-1]\geq\lambda(z)\geq 2D N^2[\cosh(z/N)-1],
\end{equation}
which is obtained using the linear response current $J=D\mathcal{A}$ with the
Einstein relation for the diffusion constant $D$ and letting $\mathcal{A}\to
0$. Thus, the bound $\eqref{eq:parabolic_eq_D}$ for the rate function of the
variable $x=X/t$ can be refined to
\begin{equation}
  \tilde h(x)\leq
  N\left[x\operatorname{arsinh}\left(\frac{x}{2DN}\right)+2DN-\sqrt{x^2+(2DN)^2}\right]
  \label{eq:cosh_eq}
\end{equation}
for unicyclic networks.

\subsection{Multicyclic networks}

A formulation of an affinity dependent bound for multicyclic networks is more
involved. In this case, the affinities of the fundamental cycles
are fixed, which means that the transition rates are constrained by relations of the 
form \eqref{eq:uniconst} for each fundamental cycle. The hyperbolic cosine bound 
for the generating function of the entropy change reads  
\begin{equation}
  \label{eq:entropy-cosh-bound-lower}
  \lambda_s(z)\geq \mep\frac{\cosh[(z+1/2)\mathcal{A}^*/n^*]-\cosh[\mathcal{A}^*/(2n^*)]}{(\mathcal{A}^*/n^*)\sinh\mathcal[\mathcal{A}^*/(2n^*)]},
\end{equation}
where the affinity $\mathcal{A}^*$ and the number of states $n^*$ corresponds to the smallest ratio $\mathcal{A}/n$ among all the cycles in in the network. 
We note that a local evaluation of the form \eqref{eq:FFbound} of the bound \eqref{eq:entropy-cosh-bound-lower}
leads to \cite{bara15a}
\begin{equation}
  F_s\geq\frac{\mathcal{A}^*}{n^*}\coth\left(\frac{\mathcal{A}^*}{2n^*}\right).
\end{equation}
This bound can be used to estimate the number of intermediate states in enzymatic 
schemes from measurements of the Fano factor in single molecule experiments, 
as discussed in \cite{bara15a}.

In order to explain how to identify $\mathcal{A}^*/n^*$ we consider
the network of states in Fig.~\ref{fig:networks}c.  We arbitrarily
choose the cycles $(1,4,2,1)$ with affinity $\mathcal{A}_1$,
$(2,4,3,2)$ with affinity $\mathcal{A}_2$, and $(1,3,4,1)$ with
affinity $\mathcal{A}_3$ as the three fundamental cycles. Any other
cycle in the network is just a composition of these fundamental
cycles; for example, the cycle $(1,4,3,2,1)$, which is marked with a
red dotted line in Fig.~\ref{fig:networks}c, with affinity
$\mathcal{A}_1+\mathcal{A}_2$ is the sum of the first and second
fundamental cycles. If the affinities are $\vec{\mathcal{A}}=(1,2,3)$
then the cycle with the smallest affinity per number of states is the
fundamental cycle $(1,4,2,1)$. Therefore, in this case
$\mathcal{A}^*=1$ and $n^*=3$.  If the affinities are
$\vec{\mathcal{A}}=(-11,12,13)$, where a negative affinity means that
the direction of the current in the cycle with affinity
$\mathcal{A}_1$ in Fig.~\ref{fig:networks}c changes from
anti-clockwise to clockwise, the cycle with minimal affinity per
number of states is $(1,4,3,2,1)$. In this case $\mathcal{A}^*=1$ and
$n^*=4$.

The basic idea behind the bound in Eq.~\eqref{eq:entropy-cosh-bound-lower} is
as follows. Given a network of states with fixed affinities, the transition
rates that lead to the smallest possible $\lambda_s(z)/J$ are those for which the
cycle with smallest $\mathcal{A}/n$ dominates the network. This cycle
dominates the network if the transition rates within the cycle are large,
transition rates to leave the cycle are small, and transition rates to return
to the cycle are large. With this choice for the transition rates the
multicyclic network is effectively a unicyclic network with affinity
$\mathcal{A}^*$ and number of states $n^*$, for which the
bound \eqref{eq:cosh-bound-lower} holds. Any other choice of rates will just
add cycles with a smaller affinity per number of states, which cannot decrease
fluctuations.

\begin{figure} 
  \centering
  \includegraphics{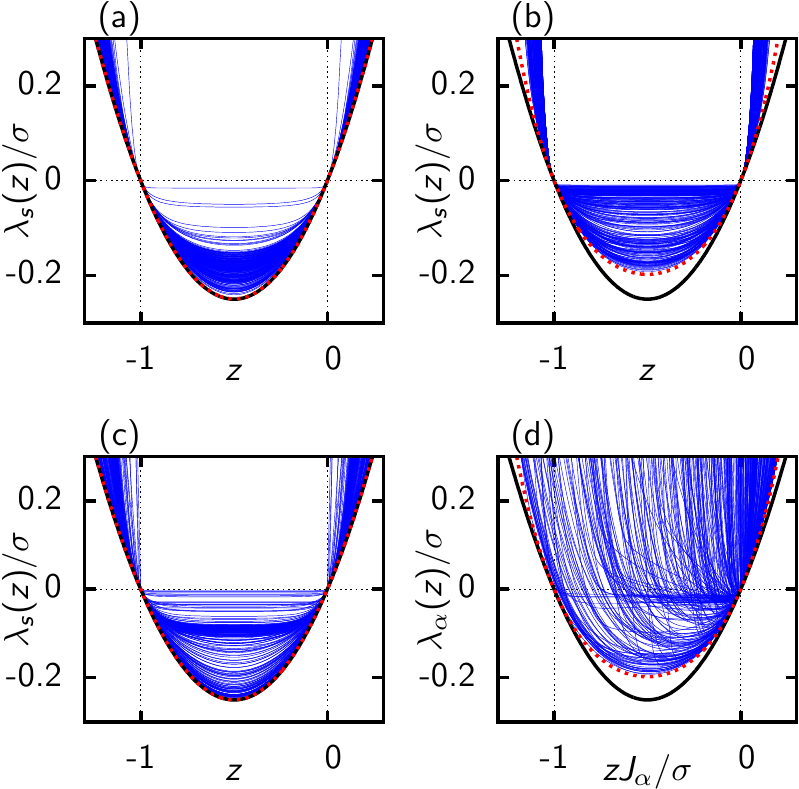}
  \caption{Generating functions for entropy change (a-c) and randomly
    selected individual currents $X_\alpha$ (d) for the network of
    Fig.~\ref{fig:networks}c. The affinities are $\mathcal{A}=(1,2,3)$
    in (a), $\mathcal{A}=(11,12,13)$ in (b) and (d), and
    $\mathcal{A}=(-11,12,13)$ in (c). The black curves correspond to the parabolic
    bound and the red dashed curves correspond to the hyperbolic cosine bound. The generating 
    functions were generated as explained in appendix~\ref{sec:numerics}.
    In panel (a) $\mathcal{A}^*/n^*= 1/3$, in panels (b) and (d) $\mathcal{A}^*/n^*= 11/3$,
     and in panel (c) $\mathcal{A}^*/n^*= 1/4$.
    For small values of $\mathcal{A}^*/n^*$, as in panels (a) and (c), the parabolic and hyperbolic 
    cosine bound are closer to each other.}
  \label{fig:bound_cosh}
\end{figure}

Our numerical evidence presented in Fig.~\ref{fig:bound_cosh}d shows that this hyperbolic cosine bound is also valid for any individual 
current $X_\alpha$ in the form
\begin{equation}
   \label{eq:mc-cosh-bound-lower}
  \lambda_\alpha(z)\geq \mep\frac{\cosh[(zJ_\alpha/\mep+1/2)\mathcal{A}^*/n^*]-\cosh[\mathcal{A}^*/(2n^*)]}{(\mathcal{A}^*/n^*)\sinh\mathcal[\mathcal{A}^*/(2n^*)]}.
\end{equation}
Hence, the hyperbolic cosine bound can be written in the more general form 
\begin{equation}
   \label{eq:mc-cosh-bound-lowervec}
  \lambda(\vec z)\geq \mep\frac{\cosh[(\vec z\cdot \vec J/\mep+1/2)\mathcal{A}^*/n^*]-\cosh[\mathcal{A}^*/(2n^*)]}{(\mathcal{A}^*/n^*)\sinh\mathcal[\mathcal{A}^*/(2n^*)]}.
\end{equation}
The full numerical evidence for this conjecture is discussed in appendix~\ref{sec:numerics}. If the cycle relevant for the bound
\eqref{eq:mc-cosh-bound-lower} has a rather small affinity per number of states, which 
is often the case in a large network of states, the bound is only slightly stronger than the parabolic bound
\eqref{eq:gf_bound_parabolic_ic}, as visible in Fig.~\ref{fig:bound_cosh}a
and~\ref{fig:bound_cosh}c. An often tighter bound for this situation is
derived in the next section. 
Our numerics indicates that an affinity dependent upper bound on the generating function in the multicyclic case does not exist. For fixed affinities, the generating function can
become arbitrarily close to the trivial bound $\lambda_s(z)<0$ for $-1<z<0$,
visible in Fig.~\ref{fig:bound_cosh}a-c. A generalization of the equilibrium
bound \eqref{eq:cosh_eq} to multicyclic networks is not directly possible,
since the identification of a cycle with minimal $\mathcal{A}/n$ becomes
ambiguous in the limit $\vec{\mathcal{A}}\to 0$.

\section{Exponential bound}
\label{sec:exponential}

A rigorous lower bound on the largest eigenvalue $\lambda(\vec z)$ of the
matrix $\mathcal{L}(\vec z)$ can be obtained from the algebraic properties
of positive matrices. Specifically, a remarkable theorem about the
largest eigenvalue $\mu$ of an arbitrary  matrix with non-negative entries $B_{ij}$ from Ellis \cite[Theorem IX.4.4]{ellis} is
\begin{equation}
  \ln \mu=\sup_{\tau_{ij}}\sum_{i,j}\tau_{ij}\ln\frac{B_{ij}\nu_i}{\tau_{ij}},
  \label{eq:ellis}
\end{equation}
where $\nu_i\equiv\sum_\ell\tau_{i\ell}$ and $0\ln 0\equiv 0$. The admissible
matrices $\tau_{ij}$ must satisfy certain normalization and symmetry
properties given in appendix~\ref{sec:expproof}. For any specific matrix $\tau_{ij}$,
Eq.~\eqref{eq:ellis} provides a lower bound on $\mu$.  

In order to apply Eq.~\eqref{eq:ellis} to the modified Markov generator $\mathcal{L}_{ij}(\vec
z)$, we construct a positive matrix 
\begin{equation}
  B_{ij}(\vec z)\equiv\delta_{ij}+\eta\mathcal{L}_{ij}(\vec z)
  \label{eq:Bellis}
\end{equation}
with a sufficiently small parameter $\eta>0$. Its largest eigenvalue is given by $1+\eta\lambda(\vec z)$. Making use of the known eigenvector of
$\mathcal{L}_{ij}(0)$, which is the stationary distribution $\pstat_i$,
we can choose $\tau_{ij}$ such that the supremum \eqref{eq:ellis} is saturated
for $\vec z=0$. As we show in appendix~\ref{sec:expproof}, fixing this choice for all 
values of $\vec z$, Eq.~\eqref{eq:ellis} yields the bound
\begin{equation}
  \lambda(\vec z)\geq(\mre^{\eta \vec z\cdot\vec J}-1)/\eta
\label{eq:ellisbound}
\end{equation}
on the generating function.

The largest possible $\eta$ in Eq.~\eqref{eq:Bellis} provides the
strongest bound.  The maximal value that still complies with the
requirement for a non-negative entries $B_{ij}$ is the inverse of the
maximal exit rate $\eta=1/\max_i(r_i)$. Extending the proof of
Eq.~\eqref{eq:ellis} in appendix~\ref{sec:expproof}, we can show that
Eq.~\eqref{eq:ellisbound} is valid for larger values of $\eta$ up to
\begin{equation}
  \eta=1/R
\end{equation}
where 
\begin{equation}
  R\equiv\sum_i\pstat_i r_i\leq\max_i r_i,
  \label{eq:defR}
\end{equation}
is the steady state activity of the network, i.e., the average number
of transitions per time interval in the steady state.  We note that a
term related to activity also appears in a fluctuation dissipation
relation for nonequilibrium steady states \cite{baie09}.

Due to the Gallavotti-Cohen symmetry \eqref{eq:GC_sym} the exponential bound can also be written as   
\begin{equation}
  \lambda(\vec z)\geq R(\mre^{(\vec{-\mathcal{A}}-\vec z)\cdot\vec J/R}-1),
  \label{eq:ellisboundsym}
\end{equation}
where we set $\eta=R^{-1}$. This bound is sharper than
\eqref{eq:ellisbound} for
$\vec z\cdot\vec J<-\vec{\mathcal{A}}\cdot\vec J/2=\mep/2$. Combining
Eqs. \eqref{eq:ellisbound} and \eqref{eq:ellisboundsym} we obtain the
exponential bound
\begin{equation}
  \lambda(\vec z)\geq R\left[\mre^{(|\mep/2+\vec z\cdot\vec
    J|-\mep/2)/R}-1\right].
  \label{eq:ellis_mc}
\end{equation}
For an individual current $X_\alpha$, the exponential bound reads  
\begin{equation}
  \lambda_\alpha(\vec z)\geq R\left[\mre^{(|\mep/2+zJ_\alpha|-\mep/2)R}-1\right].
  \label{eq:ellis_mca}
\end{equation}
The choice $\vec z= z \vec{\mathcal{A}}$ in Eq.~\eqref{eq:ellis_mc} leads to
\begin{equation}
  \lambda_s(z)\geq R\left[\mre^{(|\mep/2+ z\mep|-\mep/2)/R}-1\right]
  \label{eq:ellis_mcs}
\end{equation}
for the entropy change.

\begin{figure} 
  \centering
    \includegraphics{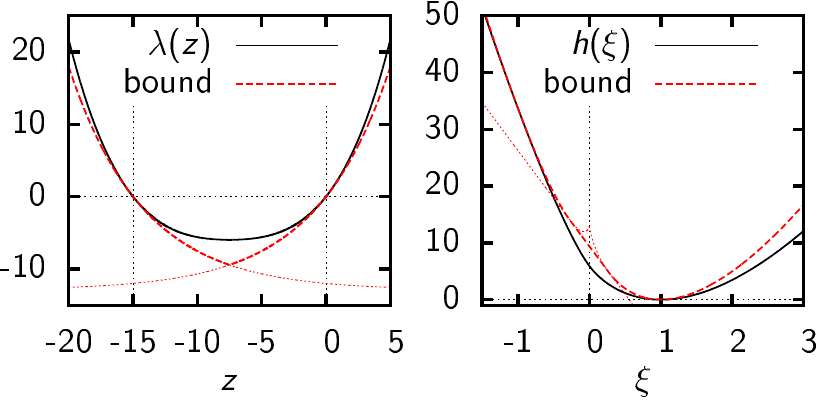}
  \caption{Generating function (left) and rate function (right) for
    a five-state unicyclic network with rates $\ln k_i^+=(3, 3, 2, 3, 2)$ and
    $\ln k_i^-=(-1, 0, -1, 0, 0)$, affinity $\mathcal{A}=15$, current
    $J\simeq2.25$, and activity $R\simeq12.9$. The functions are shown as solid
    lines and the
    exponential bound \eqref{eq:ellis_mc} as dashed lines. Analytic continuations of the piecewise
    defined functions are shown as dotted curves. }
  \label{fig:ldf_ellis}
\end{figure}

An illustration of the exponential bound \eqref{eq:ellis_mc} is
provided in Fig.~\ref{fig:ldf_ellis}. This bound is typically tighter
than the parabolic bound for far from equilibrium conditions, i.e.,
for large affinity.  For example, for a random walk on a unicyclic
network with $N$ sites, a uniform forward stepping rate $k$ and a
vanishing backward stepping rate, which implies divergent affinity,
the bound in Eq.~\eqref{eq:ellis_mc} is saturated. Specifically, in
this case the generating function is
\begin{equation}
  \lambda(z)=k\left(\mre^{z/N}-1\right),
\end{equation}
the activity is $R=k$ and the cycle current $J=k/N$. For vanishing
current at equilibrium, the exponential bound reduces to the trivial
statement $\lambda(z)\geq 0$.

Our numerics indicates that the hyperbolic cosine bound is always
tighter than the exponential bound in unicyclic networks. For
multicyclic networks the exponential bound can be
tighter. Furthermore, contrary to the hyperbolic cosine bound, the
exponential bound does not require knowledge of the topology of the
network of states, only the average entropy production and the average
activity are required.

In terms of the rate function of an individual current $X_\alpha$,
corresponding to the generating function Eq.~\eqref{eq:ellis_mca}, the
exponential bound reads
\begin{equation}
  h_\alpha(\xi)\leq
  \begin{cases}
    \frac{1}{\eta}[1+\xi-\xi\ln|\xi|]-\sigma\xi, & \xi\leq-\mre^{-\eta\sigma/2},\\
\frac{1}{\eta}[1-\xi+\xi\ln|\xi|], & \xi\geq
\mre^{-\eta\sigma/2},\\
\frac{1}{\eta}[1-\mre^{-\eta\sigma/2}]-\sigma\xi/2, & \mathrm{otherwise}.
  \end{cases}
\end{equation}
This bound on the rate function is illustrated in Fig.~\ref{fig:ldf_ellis} for a unicyclic network.

Using \eqref{eq:FFbound} in the exponential bound \eqref{eq:ellis_mc} for an individual current leads to
\begin{equation}
F_\alpha\geq J_\alpha/R.
\end{equation}
This new relation provides a lower bound on the dispersion of an
individual current, characterized by the Fano factor $F_\alpha$, in
terms of its average $J_\alpha$ and the activity $R$.

\section{Asymptotic bounds}
\label{sec:asymptotics}
\subsection{Unicyclic networks}

The asymptotic bounds discussed in the following are exact results that become tighter than all previous bounds for large values of $|z|$.
First we consider a unicyclic network with $N$ states and affinity $\mathcal{A}$. In this case, we can prove the following
bound on the generating function:
\begin{equation}
  \lambda(z)\geq J \lambda_\mathrm{ARW}(z,\mathcal{A},N)+r_{\mathrm{ARW}} -\frac{1}{N}\sum_{i=1}^N r_i,
  \label{eq:asymp_bound_uc}
\end{equation}
where $\lambda_\mathrm{ARW}(z,\mathcal{A},N)$ is defined in Eq.~\eqref{eq:arwdef},
$r_\mathrm{ARW}\equiv k^++k^-$, and
\begin{equation}
 k^\pm\equiv\left(\prod_{i=1}^Nk^\pm_{i}\right)^{1/N}. 
\end{equation}
This bound is proved in appendix~\ref{sec:asyproofuc} by comparing the weight of a trajectory in the ensemble with transition rates
$k^\pm_i$ with the weight of a trajectory in the ensemble with transition rates $k^\pm$.
Our numerics indicate that with increasing $|z|$ the difference between this bound and the actual generating function
tends to zero. This fact is quite remarkable given the exponential growth of
both functions. Unlike all other bounds presented so far, the bound
\eqref{eq:unif_rates} is not saturated at $z=0$. Only for the case of uniform
rates, i.e., $k_i^\pm=k^\pm$, the generating function \eqref{eq:gen_func_arw} saturates the bound \eqref{eq:asymp_bound_uc} globally.

\subsection{Multicyclic networks}
\begin{figure} 
  \includegraphics{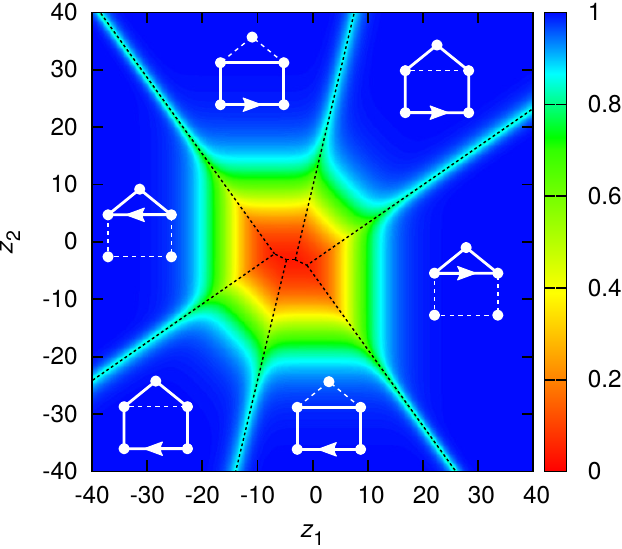}
  \caption{Asymptotic bound  for the house-shaped network with two
    fundamental cycles shown in Fig.~\ref{fig:networks}b. The color code 
    represents the ratio \eqref{eq:asymptotic_ratio} between the generating function and the
    bound. Black dashed lines indicate the borders between sectors with
    constant relevant cycles $\hat C(\vec z)$.  For each sector, the relevant
    cycle $\hat{\mathcal{C}}(\vec z)$ is shown in white. The affinity of the
    three-cycle is $\mathcal{A}_1=8$ and the affinity of the four-cycle is
    $\mathcal{A}_2=6$.}
  \label{fig:mc_asy_bound}
\end{figure}

In order to obtain an asymptotic bound also valid for multicyclic networks we define an arbitrary 
closed path $\mathcal{C}$, which is a sequence of jumps that finishes at the state it started, as 
\begin{equation}
  \mathcal{C}\equiv [i(1)\to i(2)\to\dots\to i(n_\mathcal{C})\to i(1)],
\end{equation}
where $n_\mathcal{C}$ is the length of the closed path. With this path we associate a geometric
mean of the transition rates
\begin{equation}
  \gamma_\mathcal{C}\equiv (k_{i(1),i(2)}k_{i(2),i(3)}\dots k_{i(n_\mathcal{C}),i(1)})^{1/n_\mathcal{C}}
\end{equation}
and integer winding numbers $m_\mathcal{C}^\beta$ that count how often the elementary
cycle $\beta$ is completed within the path $\mathcal{C}$.
Applying a theorem valid for arbitrary non-negative matrices \cite[Lemma 3.5.3]{bapa97}
to the matrix $\mathcal{L}_{ij}(\vec{z})+\delta_{ij}\max_\ell r_\ell$
we obtain 
\begin{equation}
  \lambda(\vec z)+\max_\ell r_\ell\geq f(\vec{z},\mathcal{C})\equiv \gamma_\mathcal{C}\exp\left(\frac{1}{n_\mathcal{C}}\sum_\beta
    m_\mathcal{C}^\beta z_\beta\right)
\label{eq:mc_asy_bound}
\end{equation}
for any closed path $\mathcal{C}$. The best bound on $\lambda(\vec z)$ in Eq.~\eqref{eq:mc_asy_bound} is obtained by choosing an optimal path $\hat{\mathcal{C}}(\vec z)$,
which in principle depends on $\vec z$, that maximizes the r.h.s of Eq.~\eqref{eq:mc_asy_bound} in the large $z$ regime.

First we consider this optimal path for the unicyclic network. In this
case, the optimal path is a single cycle in the forward direction with
$m_\mathcal{C}=1$ if $z>0$. If we consider a path $\mathcal{C}$ with
two cycles, i.e., $m_\mathcal{C}=2$, the bound remains the same as the
number of states $n_\mathcal{C}$ also doubles. If the closed path is
not a direct cycle but contains, for example, one backward jump, then
$\gamma_\mathcal{C}$ can become larger. However, such a backward jump
also makes $n_\mathcal{C}$ larger and hence, the exponent in
Eq.~\eqref{eq:mc_asy_bound} smaller. Since we are interested in the
large $z$ regime, this second effect should be dominant.  Hence, for
$z>0$ the bound \eqref{eq:mc_asy_bound} leads to
\begin{equation}
\lambda(\vec z)\ge k^+\exp(z/N)-\max_\ell r_\ell.
\end{equation}
Even though this bound is different from \eqref{eq:asymp_bound_uc},
they both predict the same exponential growth, with the same
prefactor, for large $z>0$.  The same reasoning is valid for $z<0$,
with the optimal path being a single cycle in the negative direction.

For multicyclic networks we consider the house-shaped network with
five states shown in Fig.~\ref{fig:networks}b. This network consists
of a cycle with three states and affinity $\mathcal{A}_1$ and a cycle
with four states and affinity $\mathcal{A}_2$. We choose these cycles
to be the fundamental cycles. This network also has a third cycle,
which is the cycle with five states and affinity
$\mathcal{A}_1+\mathcal{A}_2$. Given a vector $\vec =(z_1,z_2)$, the
optimal path is the cycle that maximizes the r.h.s of
Eq.~\eqref{eq:mc_asy_bound}. For large enough $|\vec z|$, this optimal
path depends only on the direction of the vector. Clearly a path that
includes other cycles will lead to a weaker bound.

A contour plot of the ratio
$f(\vec{z},\hat{\mathcal{C}}(\vec{z}))/[\lambda(\vec{z})+\max_\ell]$
for this house-shaped network is shown in
Fig.~\ref{fig:mc_asy_bound}. Remarkably, the r.h.s.  of
\eqref{eq:mc_asy_bound} captures the leading order of the asymptotics
for large $|\vec{z}|$, i.e., for large $|\vec{z}|$
\begin{equation}
  \frac{f(\vec{z},\hat{\mathcal{C}}(\vec{z}))}{\lambda(\vec{z})+\max_\ell
    r_\ell}\to 1.
\label{eq:asymptotic_ratio}
\end{equation}
Only in the lines separating regions dominated by different cycles in
Fig.~\ref{fig:mc_asy_bound} does this ratio tend to slightly lower
values. Along this line the dominant cycle is degenerate.  As shown in
appendix~\ref{sec:asyproofmc}, relation \eqref{eq:asymptotic_ratio} is
valid for any multicyclic network.  Hence, we conclude that our
asymptotic bound predicts the exponential growth of the generating
function, apart from exceptional regions in $\vec{z}$ where the
optimal cycle is degenerate.

\section{Parabolic bound in driven diffusive one dimensional systems}
\label{sec:1D}

We now consider one dimensional driven diffusive systems, which unlike
the cases considered so far have a divergent number of states $L$ in
the thermodynamic limit.  Calculating the generating function for
these systems is a major challenge that can be overcome in some cases
with the additivity principle \cite{bodi04}. In this section we
compare the parabolic bound to the cumulant generating function
obtained from this additivity principle for three examples of driven
diffusive systems, for which the validity of the additivity principle
has been verified numerically \cite{gori12,hurt10}.

\begin{figure} 
  \centering
  \sidesubfloat[]{\includegraphics{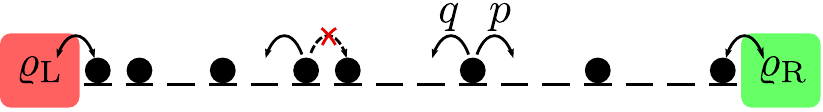}}\\
  \sidesubfloat[]{\includegraphics{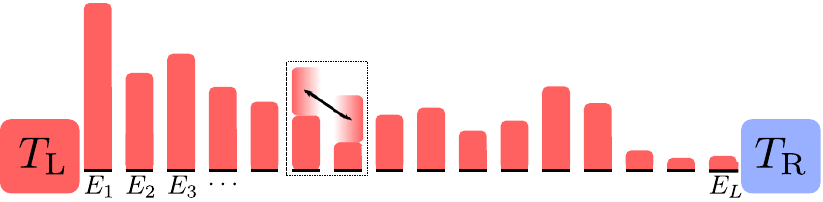}}\\
  \caption{Schematic illustrations of the WASEP (a) and the KMP (b)
    models.  For the WASEP, in the bulk the particles jump with rates
    $p\equiv1/2+\nu/(2L)$ to the right and $q\equiv1/2-\nu/(2L)$ to
    the left, where the SSEP corresponds to $\nu=0$.  At the
    boundaries particles are exchanged with the reservoirs. The model
    also has the exclusion principle, i.e., the maximum number of
    particles in a site is one. For the KMP model energy flows from a
    hot reservoir at temperature $T_L$ to a cold reservoir at
    temperature $T_R$. In the bulk a randomly chosen pair of sites
    exchange energy, which is a continuous variable, in such a way
    that the total energy is conserved. At the boundaries energy is
    exchanged with the reservoirs.  The precise rules of these models
    can be found in \cite{gori12} for the WASEP and \cite{hurt10} for
    the KMP model.  }
  \label{fig:ssep1}
\end{figure}

\begin{figure} 
  \centering
  \includegraphics{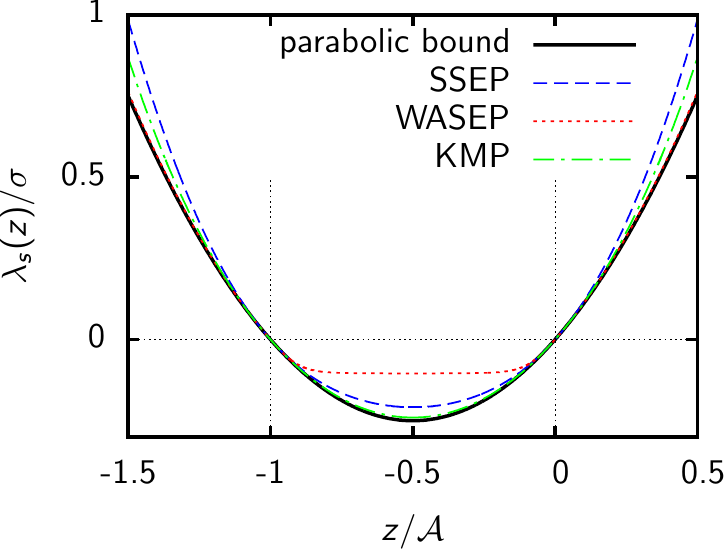}
  \caption{Comparison of the parabolic bound \eqref{eq:gf_bound_parabolic} with generating functions for driven diffusive systems. 
    For the SSEP the densities of the left and right reservoirs were chosen as $\varrho_\mathrm{L}=0.99$ and $\varrho_\mathrm{R}=0.01$. For the
    WASEP the parameters are $\nu=10$, $\varrho_\mathrm{L}=4/7$, and $\varrho_\mathrm{R}=5/18$, as in Ref.~\cite{gori12}. For the KMP model 
    the parameters are  $T_\mathrm{L}=2$ and $T_\mathrm{R}=1$, as in  Ref.~\cite{hurt10}.
    }
  \label{fig:ssep2}
\end{figure}

First we consider the WASEP and the SSEP, which is a particular case
of the WASEP. These models are illustrated in Fig.~\ref{fig:ssep1} and
their precise definition can be found in \cite{derr07}.  In the WASEP
particles flow from the left reservoir with constant density
$\varrho_\mathrm{L}$ to the right reservoir with density
$\varrho_\mathrm{R}< \varrho_\mathrm{L}$.  The current of particles in
the system is proportional to the entropy production, and the affinity
that drives the process out of equilibrium is given by
\cite{derr07,gori12}
\begin{equation}
  \mathcal{A}_{\mathrm{WASEP}}=-\ln\frac{1-\varrho_\mathrm{L}}{\varrho_\mathrm{L}}+\ln\frac{1-\varrho_\mathrm{R}}{\varrho_\mathrm{R}}+(L-1)\ln\frac{1-\nu/L}{1+\nu/L}.
\end{equation}
The weak asymmetry of the bulk rates, which scales with $1/L$,
guarantees that in the thermodynamic limit $L\to \infty$ the affinity
is finite.  In Fig.~\ref{fig:ssep2}, we have calculated the generating
function using the additivity principle for the SSEP, as explained in
\cite{derr07}, and for the WASEP, as explained in \cite{gori12}.  In
both cases the generating functions are inside the parabolic bound.

The KMP model is a driven diffusive system for the transport of energy
from a reservoir at temperature $T_\mathrm{L}$ to a reservoir at
temperature $T_\mathrm{R}< T_\mathrm{L}$, as illustrated in
Fig.~\ref{fig:ssep1}. A key feature of the KMP model is that there is
no dissipation in the bulk. The precise definition of the model can be
found in \cite{hurt10}. The heat transfer from the left to the right
reservoir is proportional to the entropy production with the affinity
given by \cite{hurt10}
\begin{equation}
  \mathcal{A}_{\mathrm{KMP}}=(T_\mathrm{R}^{-1}-T_\mathrm{L}^{-1}).
\end{equation}
The generating function for this model, which is obtained from the additivity
principle as explained in \cite{hurt10}, also satisfies the parabolic bound in
Fig.~\ref{fig:ssep2} within the finite support
$-T_\mathrm{R}^{-1}<z<T_\mathrm{L}^{-1}$ of $\lambda(z)$. As a consequence,
the rate function satisfies the corresponding parabolic bound globally.

These results demonstrate that our parabolic bound is even more universal: it
seems to be valid for these driven diffusive systems in the thermodynamic
limit, for which the number of states diverges.  We expect that the parabolic
bound is the only relevant one in the limit $L\to \infty$. The hyperbolic
cosine bound \eqref{eq:mc-cosh-bound-lower} approaches the parabolic bound for
vanishing affinity per number of states in a cycle. The exponential bound
\eqref{eq:ellisbound} degenerates with increasing activity to a linear
function, which reflects simply the convexity of the generating function.

Another interesting issue will be to explore whether the bound is still
valid in the $L\to\infty$ limit if the system undergoes a dynamical phase
transition as the KMP model in a ring-like geometry \cite{hurt11}.

\section{Conclusions}
\label{sec:conclusion}

\begin{figure} 
  \centering
  \includegraphics{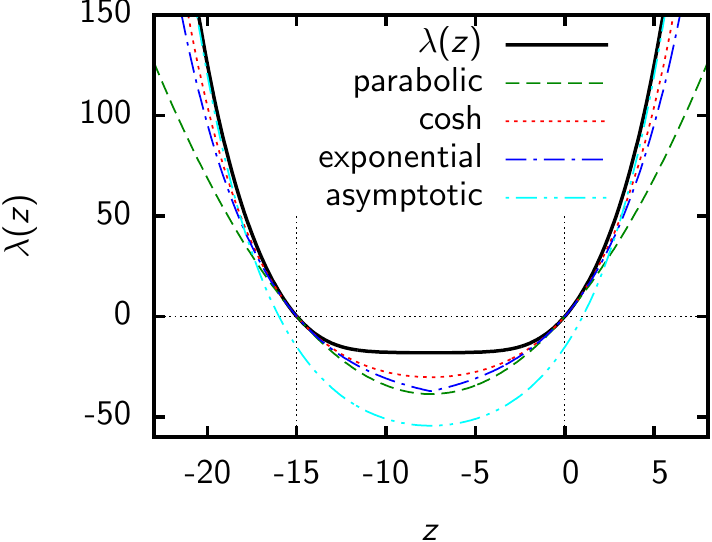}
  \caption{Summary of the four bounds for a unicyclic network with four
    states. Transition rates are $\ln k_i^+=(3, 4, 5, 4)$ and $\ln k_i^+=(0,
    -1, 1, 1)$, leading to the affinity $\mathcal{A}=15$ and the current
    $J\simeq 10.307$. }
  \label{fig:summary}
\end{figure}

We have obtained four global bounds on current fluctuations for Markov processes
in steady states summarized in table~\ref{tab:summary}. The parabolic bound
from Sec.~\ref{sec:parabolic} is the most universal result of this paper.  The
simple knowledge of the average entropy production is enough to bound the
whole range of fluctuations of any individual current. In other words, for
nonequilibrium steady states, the generating function associated with any
fluctuating current must lie inside the parabola shown in
Fig.~\ref{fig:summary}. The universality of the parabolic bound was further
confirmed by the fact that it also applies to the three driven diffusive
systems we analyzed in Sec.~\ref{sec:1D}, for which the number of states
diverges.

This parabolic bound can be saturated only close to equilibrium.  A bound that
is generally tighter than the parabolic bound, particularly if the system is
far from equilibrium, is the hyperbolic cosine bound from
Sec.~\ref{sec:cosh}. This necessarily less universal bound also requires
knowledge of the thermodynamic forces, i.e., the affinities, that drive the
process out of equilibrium and of the topology of the network of states.

The exponential bound depends on the average entropy production and on the
average number of transitions per time. In contrast to the parabolic and
hyperbolic cosine bounds that are conjectures based on extensive numerical
evidence, we have proven the exponential bound. It is typically tighter than
the parabolic bound for far from equilibrium situations. While for a unicyclic
network we observed that the hyperbolic cosine bound is always tighter than
the exponential bound, for multicyclic networks the exponential bound can be
tighter.

The fourth bound is an exact asymptotic bound that predicts the growth of the
generating function for large $z$, as illustrated in
Fig.~\ref{fig:summary}. This bound requires knowledge of the particular
transitions rates. Therefore, its importance arises in a situation where a
Markov process with all its transition rates is given but calculating the full
generating function is not possible.

Summarizing, typical and large fluctuations for any individual current in
stationary Markov processes, which are used to describe a large amount of
nonequilibrium systems ranging from enzymatic reactions to nanoscale
electronic systems, have been shown to be bounded by the average entropy
production or the average entropy production and the average
activity. Rigorous proofs of the parabolic bound and of the hyperbolic cosine
bound remain as main open technical challenges.

\textit{Note added:} A proof of the parabolic bound has recently appeared \cite{ging16}.

\begin{table}
  \centering
  \renewcommand*{\arraystretch}{3}
   \setlength{\tabcolsep}{0pt}
  \begin{tabular}{>{\centering\arraybackslash}p{2.5cm}l>{\centering\arraybackslash}p{1cm}}
\bf Parabolic  &   $\vec z\cdot \vec J\,(1+\vec z\cdot \vec J /\mep)$
 &\ \eqref{eq:lr_bound_lambda}\ \\\hline
  \bf \pbox{20cm}{Hyperbolic\\cosine}  &
$\displaystyle\mep\frac{\cosh\left[(\frac{\vec{z}\cdot\vec{J}}{\mep}+\frac{1}{2})\frac{\mathcal{A}^*}{n^*}\right]-\cosh\left[\mathcal{A}^*/(2n^*)\right]}{(\mathcal{A}^*/n^*)\sinh\left[\mathcal{A}^*/(2n^*)\right]}$
&\ \eqref{eq:mc-cosh-bound-lowervec}\ \\\hline
\bf Exponential  &  $\displaystyle{R\left[\mre^{(|\mep/2+\vec z\cdot\vec
    J|-\mep/2)/R}-1\right]}$
&\ \eqref{eq:ellis_mc}\  \\\hline
\bf Asymptotic &  $\displaystyle{-\max_\ell r_\ell+\gamma_\mathcal{C}\exp\left(\frac{1}{n_\mathcal{C}}\sum_\beta
    m_\mathcal{C}^\beta z_\beta\right)}$ &\ \eqref{eq:mc_asy_bound}\ 
  \end{tabular}
  \caption{Summary of lower bounds on the generating function. }
  \label{tab:summary}
\end{table}

\appendix

\section{Proof of the exponential bound}
\label{sec:expproof}

The theorem by Ellis \cite[Theorem IX.4.4]{ellis} can be stated as follows.
For any non-negative matrix $B_{ij}$, the associated maximum eigenvalue can be calculated as 
\begin{equation}
  \ln \mu=\sup_{\tau_{ij}}\sum_{i,j}\tau_{ij}\ln\frac{B_{ij}\nu_i}{\tau_{ij}},
  \label{eq:ellis_app}
\end{equation}
where $\nu_i\equiv\sum_k\tau_{ik}$ and $0\ln 0\equiv 0$. The admissible matrices $\tau_{ij}$ must satisfy the following properties:
\begin{enumerate}
\item Normalization,
\begin{equation}
  \sum_{i,j}\tau_{ij}=\sum_i\nu_i=1.
 \label{eq:condi1}
\end{equation}
\item Equal row- and column-sums,
  \begin{equation}
    \nu_i\equiv    \sum_\ell\tau_{i\ell}=\sum_\ell\tau_{\ell i}.
    \label{eq:condi2}
  \end{equation}
\item Non-negative with the same (or less complex) structure as $B_{ij}$, i.e.,
  \begin{equation}
    \tau_{ij}>0\Rightarrow B_{ij}>0.
  \end{equation}
\end{enumerate}

In order to apply Eq.~\eqref{eq:ellis_app} to the modified Markov generator $\mathcal{L}_{ij}(\vec{z})$, 
we consider the matrix
\begin{equation}
  B_{ij}(\vec{z})\equiv\delta_{ij}+\eta\mathcal{L}_{ij}(\vec{z}),
\end{equation}
with a sufficiently small parameter $\eta>0$. Its largest eigenvalue is $1+\eta\lambda(\vec{z})$. 
Lower bounds on the generating function $\lambda(\vec{z})$ can be obtained from \eqref{eq:ellis_app} by choosing an
appropriate matrix $\tau_{ij}$. The choice 
\begin{equation}
  \tau_{ij}=B_{ij}(0)\pstat_j
\end{equation}
saturates the bound for $\vec{z}=0$. The bound for the eigenvalue \eqref{eq:ellis_app} then reads (with $\nu_i=\pstat_i$)
\begin{equation}
  \ln[1+\eta\lambda(\vec{z})]\geq \sum_{i,j}[\delta_{ij}+\eta\mathcal{L}_{ij}(0)]\pstat_j\ln\frac{[\delta_{ij}+\eta\mathcal{L}_{ij}(\vec{z})]\pstat_i}{[\delta_{ij}+\eta\mathcal{L}_{ij}(0)]\pstat_j}.
\end{equation}
Since $\mathcal{L}_{ii}(\vec{z})=\mathcal{L}_{ii}(0)$ the logarithm vanishes 
for $i=j$ and the r.h.s. simplifies to 
\begin{align}
  \ln[1+\eta\lambda(\vec{z})]
  \geq& \sum_{i\neq
    j}\eta\mathcal{L}_{ij}(0)\pstat_j\ln\frac{\mathcal{L}_{ij}(\vec{z})\pstat_i}{\mathcal{L}_{ij}(0)\pstat_j}\\
    &=\eta\sum_{i\neq
    j}\pstat_jk_{ji}\left(\vec{z}\cdot\vec{d}_{ji}+\ln\frac{\pstat_i}{\pstat_j}\right).
\end{align}
The term proportional to $\ln(\pstat_i/\pstat_j)$ vanishes because $\pstat_i$ is the stationary distribution.  
Identifying the stationary current $\vec{J}=\sum_{i\neq  j}\pstat_i k_{ij} \vec{d}_{ij}$ we obtain
the bound
\begin{equation}
  \lambda(\vec{z})\geq\frac{1}{\eta}\left(\mre^{\eta \vec{z}\cdot\vec{J}}-1\right),
\label{eq:ellis_appbound}
\end{equation}
which is Eq.~\eqref{eq:ellisbound} in the main text.


The bound improves for larger $\eta$. The maximal value that still complies with the requirement
for positive entries $B_{ij}$ is the inverse of the maximal escape rate $\eta=1/\max_i(r_i)$. In this 
case, the proof of \eqref{eq:ellis_app} in Ref.~\cite{ellis} uses the equation
\begin{equation}
\sum_{i,j}\tau_{ij}\ln\frac{B_{ij}\nu_i}{\tau_{ij}}=\sum_{i,j}\tau_{ij}\ln\frac{B_{ij}\nu_i^0}{\tau_{ij}^0}-\sum_{i,j}\tau_{ij}\ln\frac{\tau_{ij}\,\nu_i^0}{\nu_i\,\tau_{ij}^0},
\label{eq:ellisproof1}
\end{equation}
where $\nu_i^0= \sum_j \tau_{ij}^0$. The matrix $\tau^0$ is given by
\begin{equation}
  \tau_{ij}^0=\tilde q_i B_{ij}q_j/\mu,
\label{eq:tau0}
\end{equation}
where $\tilde q$ and $q$ are left and right eigenvectors of $B$, respectively, with the
normalization $\sum_i q_i=1$ and $\sum_i \tilde q_i q_i=1$. From relations
\eqref{eq:condi1}, \eqref{eq:condi2}, and \eqref{eq:tau0} it follows that the
first sum in Eq.~\eqref{eq:ellisproof1} is
simply $\ln\mu$. The second sum with a minus sign can be shown to fulfill the inequality 
\begin{equation}
  \sum_{i,j}y_{ij}x_{ij}\ln
  x_{ij}\geq\sum_{i,j}y_{ij}(x_{ij}-1)=\sum_{i,j}(\tau_{ij}-y_{ij})=0,
\label{eq:ellisproof2}
\end{equation}
where 
\begin{equation}
  y_{ij}\equiv\frac{\nu_i\tau_{ij}^0}{\nu_i^0},\ x_{ij}\equiv \frac{\tau_{ij}\,\nu_i^0}{\nu_i\,\tau_{ij}^0}.
  \label{eq:ellisproof3}
\end{equation}
Equality in Eq.~\eqref{eq:ellisproof2} is achieved for $\tau_{ij}=\tau_{ij}^0$, which with Eq.~\eqref{eq:ellisproof2} 
provides a proof of \eqref{eq:ellis_app}.

Actually, we can show that the bound \eqref{eq:ellis_appbound} is
valid even for larger values of $\eta$ up to $\eta=R^{-1}$, where $R$
is the average activity in Eq.~\eqref{eq:defR}. In this case, the
diagonal elements $B_{ii}$ can be negative. This property can affect
Eq.~\eqref{eq:ellisproof2}, which requires that $x_{ij}\ge0$ and
$y_{ij}\ge0$. For
$B_{ij}(z)=\delta_{ij}+\eta\mathcal{L}^\alpha_{ij}(z)$, if
\begin{equation}
1+\eta\lambda_\alpha(z)>0
\label{eq:condcond}
\end{equation}
then all $x_{ij}$ and $y_{ij}$ in Eq.~\eqref{eq:ellisproof3} for
$i\neq j$ are non-negative.  The inequality \eqref{eq:ellisproof2} is
then valid for $i\neq j$ with condition \eqref{eq:condcond}.  For the
diagonal terms we can write
\begin{align}
\sum_i y_{ii} x_{ii}\ln x_{ii}&=\sum_i \pstat_i (1-\eta
r_i)\ln[1+\eta\lambda_\alpha(z)]\nonumber\\
&\geq \sum_i\pstat_i(1-\eta r_i)\frac{\eta\lambda_\alpha(z)}{1+\eta\lambda_\alpha(z)},
\end{align}
where the inequality is valid for $\eta\le R^{-1}$ and if condition \eqref{eq:condcond} is fulfilled. Therefore, if condition \eqref{eq:condcond} holds then the inequality \eqref{eq:ellisproof3} 
is valid for $\eta\le 1/R$. 

Since this condition \eqref{eq:condcond} is valid for $z=0$, it must
also be valid for some finite range in $z$. Using a simple
self-consistency check, we can even prove that this ``finite range''
must in fact always be infinite. Assume the function
$1+\lambda_\alpha(z)/R$ crosses zero at some value $z=z^*$. Then the
condition \eqref{eq:condcond} is violated for $z=z^*-\delta z$. On the
other side, for $z=z^*+\delta z$, the condition is still satisfied and
we find
\begin{equation}
  1+\lambda_\alpha(z^*+\delta z)/R\geq\mre^{(z^*+\delta z)J_\alpha/R}>0,
\end{equation}
which contradicts the continuity of $\lambda_\alpha(z)$.

\section{Numerical verification}
\label{sec:numerics}

The numerical verification of the conjectured bounds was performed on large
sets of networks with different transition rates. The corresponding generating
functions are given by the largest eigenvalues of the matrices
$\mathcal{L}(z)$. We calculated these eigenvalues using standard numerical algorithms.
The stationary distributions $\pstat_i$, which are computed as the eigenvector for $z=0$, are
used to evaluate the steady state currents that appear in the bounds. The precise procedures 
are described below.

\subsection{Unicyclic networks}

For unicyclic networks with $N$ states it is convenient to parametrize the
transition rates \eqref{eq:unif_rates} as
\begin{equation}
  k_i^\pm=\exp(\phi_i\pm\theta_i\mathcal{A}/2).
\end{equation}
The global time scale can be fixed by requiring $\sum_i\phi_i=0$. Thus
we avoid numerical instabilities due to extremely large or small
matrix entries. Moreover, in order to sample cycles with predefined
affinity $\mathcal{A}$, we require $\sum_i\theta_i=1$.  We generate
vectors $\phi'_i$ and $\theta'_i$ of $N$ independent and identically
distributed random numbers.  The above constraints are satisfied by
setting $\phi_i=\phi'_i-\overline{\phi'}$ and
$\theta_i=\theta'_i/\overline{\theta'}$, where the overbar denotes the
average of the vectors elements within the realization. Samples where
at least one of the $|\theta_i|$ is above a certain value were
discarded in order to avoid numerical instabilities. The cutoff value
$1$ turns out to be suitable for this purpose. Since the transition
rates associated with the discarded samples are extremely non-uniform,
the corresponding generating function lies close to the (proven) upper
bound anyway.

For the plots shown in Fig.~\ref{fig:bounds_uc}, $\phi'_i$ and
$\theta'_i$ were drawn from uniform distributions with $0<\phi_i'<4$
and $-0.5<\theta_i'<0.5$, respectively. The generating functions were
calculated for a total of $10\;000$ samples for each affinity, of
which only the first $100$ are shown in the figure. For cycles with
many states and high affinity, it is virtually impossible to cover the
whole area between the upper and the lower bound using a single,
simple distribution for $\phi'_i$ and $\theta'_i$. In
Fig.~\ref{fig:bounds_uc_highA}, we show several families of sample
generating functions where the rates are drawn from different
statistical ensembles. Specifically, $\phi'_i$ and $\theta'_i$ are
both Gaussian random variables with mean 1 (which is irrelevant for
$\phi_i$) and standard deviations (SD) reaching from $0.01$ to $2$.

In principle the lengths sub-steps $d_{i,i+i}$ can be distributed
arbitrarily among the edges of the network. In most cases, the choice
$d_{i,i+i}\propto\ln(k_i^+/k_i^-)$ avoids numerical instabilities. In
order to pre-assess the range of $z$ we can make use of the proven
asymptotic bound \eqref{eq:asymp_bound_uc}: the validity of the
hyperbolic cosine bound has to be checked only in the finite range
where it is weaker than the asymptotic bound. In all cases the
hyperbolic cosine bound \eqref{eq:cosh-bound-lower} is
satisfied. Since the hyperbolic cosine bound implies the parabolic
bound, this numerical evidence also allows us to conjecture the parabolic
bound for unicyclic networks.

\subsection{Multicyclic networks}

The bounds relevant for a numerical test for multicyclic networks are the
parabolic bound and the hyperbolic cosine bound. They exist in the
formulation for entropy change [Eqs. \eqref{eq:gf_bound_parabolic_entropy}
and \eqref{eq:entropy-cosh-bound-lower}] and for arbitrary individual currents [Eqs. \eqref{eq:gf_bound_parabolic_ic}
and \eqref{eq:mc-cosh-bound-lower}]. The former type can be checked by setting the
matrix of increments in Eq.~\eqref{eq:defL} to $d^s_{ij}=\ln(k_{ij}/k_{ji})$, for the latter we use
anti-symmetrized Gaussian random matrices for $d_{ij}^\alpha$.

We have performed two types of tests. 
The first type relies on random rate matrices $k_{ij}$ of dimension
$N\times N$, each of them corresponding to a fully connected network with $N$
states. The rates were generated according to
\begin{equation}
  k_{ij}=\exp[a(\phi_{ij}+\phi_{ji})/2+b\theta_{ij}],
\end{equation}
where $\phi_{ij}$ and $\theta_{ij}$ are independent Gaussian random
numbers with zero mean and variance 1. The parameters $a$ and $b$ can
be used to tune the properties of the network. While small values of
$a$ simulate fully connected networks, larger values of $a$ typically
suppress some of the transitions, so that the generated matrices
effectively correspond to partially connected networks with random
topology. The parameter $b$ introduces an asymmetry in the
transitions, that drives the system out of equilibrium. For smaller
values of $b$ the generating functions lie closer to the parabolic
bound. For Fig.~\ref{fig:bound_parabolic} we have calculated 300
generating functions for each $N=4$ and $N=6$ with $a=5$ and $b=2$. In
a more extensive computation, we have checked the parabolic bound for
a total of $10^7$ generating functions with $a$ ranging from 0 to 5,
$b$ ranging from 0.01 to 5, and $N$ ranging from $4$ to $50$. The
hyperbolic cosine bound could be checked only up to $N=8$, for larger
networks the determination of the relevant cycle in
\eqref{eq:mc-cosh-bound-lower} becomes numerically expensive.

The second type of test applies to small networks with given topology, where
the fundamental cycles can be identified by hand. The affinities of these
cycles can be fixed, so that the bounds \eqref{eq:entropy-cosh-bound-lower}
and \eqref{eq:mc-cosh-bound-lower} depend only on the steady state
currents. For example, for the network shown in Fig.~\ref{fig:networks}c,
random rates were assigned to most of the transition. The random numbers were
generated such that $\ln k_{ij}$ were Gaussian random numbers with standard
deviations ranging from 0.01 to 3 and, at first, with zero mean. Only the
three forward transition rates in the cycle (1,2,3,1) were determined
algebraically from the other transition rates and the constraints from the
fixed cycle affinities $\mathcal{A}_1$, $\mathcal{A}_2$ and $\mathcal{A}_3$.
Typically, the generating functions obtained via this procedure are quite far
from the hyperbolic cosine bound. In order to test also more critical cases,
we have added a bias $\pm\mathcal{A}^*/(2n^*)$ to the logarithms of the
forward (+) and backward (-) transition rates constituting the relevant cycle for the
hyperbolic cosine bound. Moreover, the rates for transitions exiting the
relevant cycle were gradually lowered by several orders of magnitude. For the
plots in Fig.~\ref{fig:bound_cosh}, we have used attenuations of these rates
between $\mre^0$ and $\mre^{-25}$. Similar tests were performed for a large
varieties of networks (as the networks shown in Fig. 2 of the supplementary material of
Ref.~\cite{bara15}).

\section{Proof of the upper bound on the generating function for unicyclic
  networks}
\label{sec:upperbound}

For unicyclic networks with $N$ states the tilted Markov generator has the tridiagonal shape
\begin{equation}
\mathcal{L}(z)=
\left(  \begin{matrix}
    -r_1 & \tilde k_{21} & 0 & 0 & \dots & \tilde k_{N1}\\
\tilde k_{12} &    -r_2 & \tilde k_{32} & 0 & \dots& 0 \\
0 & \tilde k_{23} &    -r_3 & \tilde k_{43} &  \ddots &0 \\
0 & 0 & \tilde k_{34} &    -r_4 & \ddots & \vdots  \\
\vdots & \vdots &\ddots &\ddots &\ddots &\tilde k_{N,N-1} \\
\tilde k_{1N} & 0 & 0 & \dots &   \tilde k_{N-1,N} & -r_N   \\
  \end{matrix}\right)
\label{eq:uc_matrix}
\end{equation}
with $\tilde k_{ij}\equiv k_{ij}\mre^{z d_{ij}}$. The displacements $d_{ij}$
must satisfy $d_{ij}=-d_{ji}$. If the observable of interest is the number of
turnovers, the displacements must add up to the cycle affinity
$d_{12}+d_{23}+\dots+d_{N1}=1$. It should be kept in mind that any statistical
quantity in the long time limit (in particular the generating
function and the rate function) do not depend on the specific
choice of the individual $d_{ij}$.

\begin{figure}
  \centering
  \includegraphics{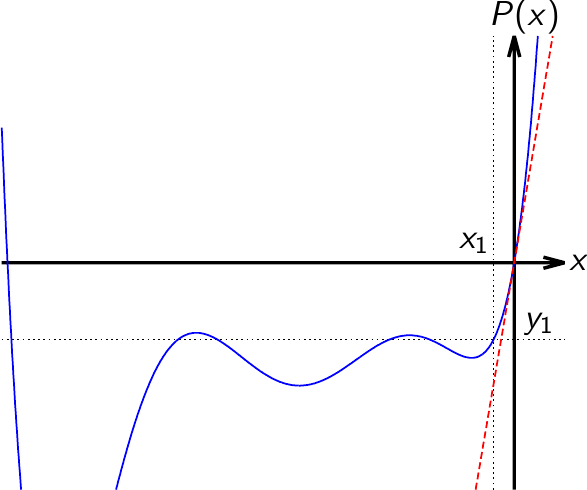}
  \caption{The polynomial $P(x)$ for a generic unicyclic network with $N=6$
    states. The tangent at $x=0$ and the values $y=y_1$ and $x=x_1$ are shown
    as dashed lines. }
  \label{fig:char_poly}
\end{figure}

The characteristic polynomial associated with the matrix \eqref{eq:uc_matrix} reads
\begin{align}
  \chi(z,x)&\equiv\det(\mathcal{L}(z)-x\mathbf{1}_N)\nonumber\\
&=\sum_\pi
  (-1)^\pi \prod_{i=1}^N(\mathcal{L}_{i\pi(i)}(z)-x\delta_{i\pi(i)}),
\label{eq:charf}
\end{align}
where the sum runs over all permutations $\pi$ of the indices
$i=1,\dots,N$ and $\mathbf{1}_N$ is the $N\times N$ identity matrix. We identify $0\equiv
N$ and $N+1\equiv 1$ for the indices of matrix entries. There are two types of
terms in \eqref{eq:charf} that contain a specific rate $\tilde k_{i+1,i}$: the contribution from the next row $i+1$ can either be $\tilde k_{i,i+1}$
or $\tilde k_{i+2,i+1}$. For the former type the $z$-dependence cancels
out due to $d_{i+1,i}=-d_{i,i+1}$ and we end up with the constant factor
$\tilde k_{i+1,i}\tilde k_{i,i+1}=k_{i+1,i}k_{i,i+1}$.
Terms of the latter type must also contain $\tilde k_{i,i-1}$ as the only
possible contribution from the previous column $i-1$. Iteratively, we see that
there can be only one term of this type, namely the one that contains all
forward transitions
\begin{equation}
  \tilde k_{12}  \tilde k_{23} \dots \tilde k_{N1}=   k_{12}   k_{23} \dots
  k_{N1}\mre^{z}\equiv \varGamma^+\mre^{z}.\label{eq:diag1}
\end{equation}
An analogous argument can be set up for the lower off-diagonal of the matrix
with the $z$-dependent term
\begin{align}
  \tilde k_{21}  \tilde k_{32} \dots \tilde k_{1N}=   k_{21}   k_{32} \dots
  k_{1N}\mre^{-z}\nonumber\\
\equiv \varGamma^-\mre^{-z}=\varGamma^+\mre^{-(z+\mathcal{A})}.\label{eq:diag2}
\end{align}
All other terms in the determinant \eqref{eq:charf} do not depend on $z$
and we can write
\begin{equation}
  \chi(z,x)=(-1)^{N+1}[\varGamma^+\mre^{z}+\varGamma^-\mre^{-z}-(\varGamma^++\varGamma^-)-P(x)]
\end{equation}
with some polynomial $P(x)$ that is independent of $z$ and the specific choice
of the $d_{ij}$. The alternating prefactor is due to the fact that
the permutations associated with the terms \eqref{eq:diag1} and
\eqref{eq:diag2} are either odd or even, depending on the number of states
$N$. The generating function is thus given by
\begin{align}
  \lambda(z)&=P^{-1}\left(\varGamma^+\mre^{z}+\varGamma^-\mre^{-z}-\varGamma^+-\varGamma^-\right)
\nonumber\\
&=P^{-1}\left(2\sqrt{\varGamma^+\varGamma^-}[\cosh(z+\mathcal{A}/2)-\cosh(\mathcal{A}/2)]\right),\label{eq:alpha_poly}
\end{align}
where the function $P^{-1}(y)$ returns the root of the polynomial $P(x)-y$
that has the largest real part. Due to the Perron-Frobenius theorem, this root
must be real for all arguments occurring in \eqref{eq:alpha_poly}, i.e., for all
$y\geq y_1\equiv 2\sqrt{\varGamma^+\varGamma^-}[1-\cosh(\mathcal{A}/2)]$. 
The root associated with the minimal argument $y_1$ is $x_1\equiv P^{-1}(y_1)=\min_z\lambda(z)=\lambda(-\mathcal{A}/2)$.
Obviously, the polynomial $P(x)$ (see Fig.~\ref{fig:char_poly}) has the properties $P(0)=\chi(0,0)=0$ and
\begin{align}
  \lim_{x\to\infty}P(x)&=(-1)^{N}\lim_{x\to\infty}\chi(z,x)\nonumber\\
&=\lim_{x\to\infty}(-1)^{N}\det(-x\mathbf{1}_N)=+\infty.
\end{align}
Since the matrix $\mathcal{L}(-\mathcal{A}/2)$ can be brought to a symmetric
form by choosing $d_{ij}=\ln(k_{ij}/k_{ji})/\mathcal{A}$, the corresponding
characteristic polynomial $P(x)-y_1$ has only real roots $x_i$ with $x_1$ denoting the
largest one. The second derivative of $P(x)$ is
\begin{align}
  P''(x)=\frac{\mrd^2}{\mrd
    x^2}\left[y_1+\prod_{i=1}^N(x-x_i)\right]=\sum_{i=1}^N\sum\limits_{\substack{j=1\\j\neq
    i}}^N\prod\limits_{\substack{\ell=1\\i\neq\ell\neq
    j}}^N(x-x_\ell).
\end{align}
For $x>x_1$ this expression is positive so that $P(x)$ is convex. As a
consequence, the inverted function $P^{-1}(y)$ is concave for the relevant
arguments $y>y_1$. Hence it satisfies
\begin{equation}
  P^{-1}(y)\leq (P^{-1})'(0)\,y
\end{equation}
with equality for $y=0$. This relation leads to the upper bound 
\begin{equation}
  \lambda(z)\leq 2\sqrt{\varGamma^+\varGamma^-}(P^{-1})'(0)\left[\cosh(z+\mathcal{A}/2)-\cosh(\mathcal{A}/2)\right]
\end{equation}
holding with equality for $z=0$. 
The prefactor in this bound is equal to the one in
Eq.~\eqref{eq:cosh-bound-upper}, as can be seen by calculating steady state
current from Eq.~\eqref{eq:alpha_poly},
\begin{equation}
  J=\lambda'(z)=2\sqrt{\varGamma^+\varGamma^-}(P^{-1})'(0)\sinh(\mathcal{A}/2).
\end{equation}

\begin{widetext}

\section{Proof of the asymptotic bound for unicyclic networks}
\label{sec:asyproofuc}

First  we restrict to a unicyclic network with $N$ states, affinity
$\mathcal{A}$ and transition rates
\begin{equation}
  k_{ij}\equiv\delta_{i,i+1}k_i^++\delta_{i+1,i}k_{i+1}^-.
\end{equation}
A stochastic path $n(\tau)$ is defined by the sequence of jumps $n_\ell\to
n_{\ell+1}$ between adjacent states that occur at times $\tau_\ell$. The weight of this path is
given by
\begin{equation}
  \mathcal{P}[n(\tau)]=\prod_\ell k_{n_\ell n_{\ell+1}}\exp[-r_{n_\ell}(\tau_{\ell+1}-\tau_\ell)],
\end{equation}
where the sum runs over all jumps. The weight of a path with modified transition rates $\tilde k_{ij}$ reads
\begin{equation}
  \tilde{\mathcal{P}}[n(\tau)]=\prod_\ell \tilde k_{n_\ell n_{\ell+1}}\exp[-\tilde r_{n_\ell}(\tau_{\ell+1}-\tau_\ell)].
\end{equation}
For these modified transition rates we choose an asymmetric random walk, i.e.,
\begin{equation}
  \tilde k_{ij}\equiv\delta_{i,i+1}k^++\delta_{i+1,i}k^-
\end{equation}
with
\begin{equation}
  k^\pm\equiv\left(\prod_{i=1}^N k_i^\pm\right)^{1/N},
\label{eq:gammas}
\end{equation}
which leads to $\tilde r\equiv k^++k^-$. Ensemble averages using the path weight $\tilde{\mathcal{P}}[n(\tau)]$ are
denoted as $\mean{\dots}_\mathrm{ARW}$. The generating function can be rewritten as 
\begin{align}
  \lambda(z)&=\frac{1}{t}\ln\mean{\mre^{zX[n(\tau)]}}=\frac{1}{t}\ln\int \mathcal{D}_{n(\tau)} \mre^{zX[n(\tau)]}\frac{\mathcal{P}[n(\tau)]}{\tilde{\mathcal{P}}[n(\tau)]}\tilde p[n(\tau)]\nonumber\\
&=\frac{1}{t}\ln\mean{\mre^{zX[n(\tau)]}\prod_{i=1}^N\left(\frac{k_{i}^+}{\gamma^+}\right)^{m_i^+}\left(\frac{k_{i}^-}{\gamma^-}\right)^{m_i^-}\mre^{-(r_i-\tilde
  r_i)\mathcal{T}_i}}_\mathrm{ARW},
  \label{eq:aaa}
\end{align}
where the integration in the first line is over all stochastic trajectories. 
The path dependent variables $m_i^\pm$ count the jumps out of state $i$ in
forward or backward direction and $\mathcal{T}_i$ is the total sojourn time in
state $i$. These variables are identically distributed in the ARW-ensemble.

The probability $\tilde P(X)$ is the probability that the fluctuating current is $X$ 
in the ARW-ensemble. It is the sum of the weight of all trajectories for which the current 
is $X$. Using this $\tilde P(X)$ Eq.~\eqref{eq:aaa} can be written as 
\begin{align}
  \lambda(z)&=\frac{1}{t}\ln\sum_X\tilde
  p(X)\mre^{zX}\mean{\exp\left[\sum_{i=1}^Nm_i^+\ln(k_i^+/\gamma^+)+\sum_{i=1}^Nm_i^-\ln(k_i^-/\gamma^-)-\sum_{i=1}^N(r_i-\tilde
    r)\mathcal{T}_i\right]\Bigg|X}_\mathrm{ARW}\nonumber\\
&\geq \frac{1}{t}\ln\sum_X\tilde
  p(X)\mre^{zX}\exp\left[\sum_{i=1}^N\mean{m_i^+|X}_\mathrm{ARW}\ln(k_i^+/\gamma^+)+\sum_{i=1}^N\mean{m_i^-|X}_\mathrm{ARW}\ln(k_i^-/\gamma^-)-\sum_{i=1}^N(r_i-\tilde
    r)t/N\right],
\end{align}
where the conditioned average in the first line represents a functional integration over all trajectories with fluctuating current equal to $X$ and we used Jensen's inequality 
from the first to the second line. Due to \eqref{eq:gammas} the terms with the logarithms vanish, 
leading to the final result in Eq.~\eqref{eq:asymp_bound_uc}.

\section{Proof of the asymptotic limit (\ref{eq:asymptotic_ratio}) for
  multicyclic networks}
\label{sec:asyproofmc}

For general Markov generators $\mathcal{L}(\vec z)$, as defined in
Eq.~\eqref{eq:defL}, the determinant \eqref{eq:charf} can be written 
as
\begin{equation}
  \label{eq:charf_mc}
  0=\chi(\vec{z},\lambda(\vec{z}))=\prod_{i=1}^N[-r_i-\lambda(\vec{z})]+\sum_\mathcal{C}(-1)^\mathcal{C}\gamma_\mathcal{C}^{n_\mathcal{C}}\mre^{\vec{m}_\mathcal{C}\cdot\vec{z}}\prod_{j\notin\mathcal{C}}[-r_j-\lambda(\vec{z})],
\end{equation}
where the sum runs over all combinations of disjoint cycles in the underlying
network and $(-1)^\mathcal{C}$ denotes the sign of the corresponding
permutations in the determinant. For each $\mathcal{C}$, the quantities
$n_\mathcal{C}$, $\gamma_\mathcal{C}$ and $\vec{m}_\mathcal{C}$ are defined as
for the individual cycles in Sec.~\ref{sec:asymptotics} of the main text. Dividing Eq. \eqref{eq:charf_mc}  by $\lambda(\vec{z})^N$ leads to
\begin{equation}
  \label{eq:charf_mc2}
  0=\prod_{i=1}^N[-r_i/\lambda(\vec{z})-1]+\sum_\mathcal{C}(-1)^\mathcal{C}f(\vec
  z,\mathcal{C})^{n_\mathcal{C}}\lambda(\vec
  z)^{-n_\mathcal{C}}\prod_{j\notin\mathcal{C}}[-r_j/\lambda(\vec{z})-1],
\end{equation}
where $f(\vec{z},\mathcal{C})$ is defined in Eq.~\eqref{eq:mc_asy_bound}.
We now analyze the limit $|\vec z|\to\infty$ with the direction
$\vec{z}/|\vec{z}|$ kept fixed. Making use of the (already proven) lower bound
\eqref{eq:mc_asy_bound} with the optimal path
$\hat{\mathcal{C}}\equiv\hat{\mathcal{C}}(\vec z)$, we see that $r_i/\lambda(\vec z)$ and 
the terms with $(\vec{m}_\mathcal{C}\cdot\vec{z})/{n_\mathcal{C}}<\vec{m}_{\hat{\mathcal{C}}} \cdot\vec{z}/{n_{\hat{\mathcal{C}}}}$
vanish in this limit. Provided that the optimal cycle is unique, we are left
with
\begin{equation}
  0=(-1)^N+\lim_{|\vec z|\to\infty}(-1)^{1+n_{\hat{\mathcal{C}}}}f(\vec
  z,{\hat{\mathcal{C}}})^{n_{\hat{\mathcal{C}}}}\lambda(\vec{z})^{-n_{\hat{\mathcal{C}}}}(-1)^{N-n_{\hat{\mathcal{C}}}},
\end{equation}
which leads to
\begin{equation}
  \lim_{|\vec z|\to\infty}\frac{f(\vec
  z,{\hat{\mathcal{C}}})}{\lambda(\vec z)}=1.
\end{equation}
In Eq.~\eqref{eq:asymptotic_ratio}, the constant $\max_\ell r_\ell$ is added to
the denominator without harm, in order to make the ratio positive everywhere.
The essential ingredient in this proof is the uniqueness of the optimal cycle
$\hat{\mathcal{C}}$. Only in peculiar regions the vector $\vec z$ leads to
more than one cycle with the same value of $\vec{m}_\mathcal{C}\cdot\vec{z}/n_\mathcal{C}$.
For example, these regions show up in Fig.~\ref{fig:mc_asy_bound} as the lines along which the ratio
\eqref{eq:asymptotic_ratio} differs from 1. 

\end{widetext}

\end{document}